\documentclass{article}
\usepackage[utf8]{inputenc}

\usepackage{nida-macros}
\usepackage{ulem} %

\definecolor{mycolor1}{rgb}{0.00000,0.44700,0.74100}

\usepackage{combelow}

\usepackage[affil-it]{authblk}

\title{Newton-Okounkov bodies of chemical reaction systems}
\author[1]{Nida Obatake} 
\author[2]{Elise Walker}
\affil[1]{Institute for Defense Analyses, Center for Communications Research La Jolla}
\affil[2]{Department of Mathematics, Texas A\&M University}

\date{February 8, 2022}

\begin{document}

\maketitle

\begin{abstract}
    \noindent 
    Despite their noted potential in polynomial-system solving, there are few concrete examples of Newton-Okounkov bodies 
    arising from applications. 
    Accordingly, in this paper, we introduce a new application of Newton-Okounkov body theory to the study of chemical reaction networks, and compute several examples. 
    An important invariant of 
    a chemical reaction network is its maximum number of positive steady states%
    , which is realized as the maximum number of positive real roots of a parametrized polynomial system. 
    Here, we introduce a new upper bound on this number, namely the 
    `Newton-Okounkov body bound' of 
    a chemical reaction network. 
    Through explicit examples, we show that the 
    Newton-Okounkov body 
    bound of 
    a network gives a good upper bound on its maximum number of positive steady states. 
    We also compare this 
    Newton-Okounkov body bound 
    to a related upper bound, namely the mixed volume of a chemical reaction network, and find that it often achieves better bounds.
\end{abstract}

\section{Introduction}
    A Newton-Okounkov body associated to a projective variety is a convex body 
    that encodes important information about the variety. 
    Newton-Okounkov bodies associated to projective varieties can be thought of as a vast generalization of Newton polytopes associated to projective toric varieties. 
    One important parallel 
    is that the volume of the convex body and is directly related to the degree of the associated projective variety, which then conveys the number of solutions to a general polynomial system.

    In contrast to Newton polytopes of projective toric varieties, 
    Newton-Okounkov bodies associated to arbitrary projective varieties 
    require an involved construction. 
    Part of the contribution of this paper is a minimal, yet explicit explanation of the setting and construction of a Newton-Okounkov body. 
    We refer the interested reader to the foundational works~\cite{kk-nob-1, lm-nob} for technical details of Newton-Okounkov body theory.  

    With an eye towards applications, our focus is on the 
    use 
    of Newton-Okounkov bodies in intersection theory, specifically to compute bounds 
    on the number of isolated solutions to polynomial systems. 
    The starting data is a finite dimensional vector subspace of rational functions $V$ contained in the coordinate ring of 
    a $d$-dimensional complex algebraic variety $X$. 
    Given such a vector space $V$, Kaveh and Khovanskii~\cite{kaveh-khovanskii-first-def-biii-2010} defined the \textit{(birationally-invariant)~self-intersection index} of $V$ to be the number of ``effective'' solutions in $X$ to a system of $d$ general equations $f_1=\dots=f_d=0$, where each $f_i$ is a general rational function in $V$. 
    Here we use ``effective'' solution to mean that the intersection index counts only solutions in the \textit{smooth locus} of $X$, and outside the \textit{base locus} (see Definition~\ref{def:effective-solution}). 
    
    As an example, 
    the classical Kushnirenko Theorem (Proposition~\ref{prop:kushnirenko}) computes the self-intersection index of a subspace $V$ of Laurent polynomials. 
    More explicitly, consider a subspace $V \subset \C[x_1^{\pm 1},\dots,x_d^{\pm 1}]$ spanned by Laurent monomials with 
    support 
    contained in some finite subset $A\subset \Z^d$. %
    The self-intersection index of $V$ is equal to the normalized Euclidean volume of the common Newton polytope $\Delta:=\conv{A}$. 
    In other words, a general system of $d$ 
    Laurent  
    polynomials in $V$ 
    with Newton polytope $\Delta=\conv{A}$ 
    has $d!\,\vol{\Delta}$ effective solutions (all in $\C^*$). 
    The Newton polytope is equal to the Newton-Okounkov body in this setting.
    
    Generalizing from vector spaces spanned by monomials, Kaveh and Khovanskii~\cite{kk-nob-1} subsequently gave a formula for the self-intersection index for subspaces $V$ spanned by rational functions. 
    That is, the number of effective solutions to a general system of $d$ equations 
    drawn from a 
    subspace of rational functions $V$  
    is proportional to  
    the volume of an associated Newton-Okounkov body $\Delta$%
    ; see Proposition~\ref{prop:kk-nob-bound}.

    The main challenge of Kaveh and Khovanskii's formula is the computation of an associated Newton-Okounkov body. 
    Such convex bodies are not necessarily polytopes, nor are the associated semigroups even finitely generated. 
    However, the Newton-Okounkov body will be a polytope if there exists an associated \textit{finite Khovanskii basis}~\cite{anderson-nob}. 
    The formalization of Khovanskii bases, which generalize sagbi bases~\cite{robbiano1990subalgebra,sturmfels-algs-in-invariant-theory}, is attributed to Kaveh and Manon, who described Khovanskii bases as the computational and algorithmic side of the theory of Newton-Okounkov bodies~\cite{kaveh-manon}. 
    This computational side of Newton-Okounkov bodies has been explored in 
    polynomial-system solving, such as in~\cite{BurrSottileWalker,duff-hein-sottile}. 
    While 
    these works focused on computing all solutions to a given polynomial system, our work is concerned with 
    computing bounds on the number of solutions to polynomial systems arising from applications.
    Specifically, our application comes from counting \emph{steady states} of a \emph{chemical reaction network}.   
    Concretely, our focus is on 
    computing a 
    self-intersection index associated to a given chemical reaction network, and assessing the resulting bound on the maximum number of the network's steady states.     

    A chemical reaction network is a model of the interactions of chemical species. 
    Under 
    a 
    classical assumption {(\emph{mass-action kinetics})},  
    the 
    network's
    dynamics
    are governed by an \emph{autonomous system of parametrized polynomial ordinary differential equations} -- see equation~\eqref{eq:ode}.
    Then, finding \emph{steady states} of a chemical reaction network amounts to solving a system of (parametrized) polynomial equations.
    The \emph{capacity} for multiple steady states (called \emph{multistationarity}) has been studied extensively in recent years, with algebraic-geometric methods at the forefront of new results~\cite{CFMW, ME_entrapped, DPST, FeinOsc, Giaroli-Bihan-Dickenstein, mss-review,signs}. 
    After deciding multistationarity, actually determining an arbitrary reaction network's maximum number of observable steady states is still an important open problem. 
    One strategy is to produce good upper bounds on the maximum number of observable steady states. 
    
    Accordingly, Obatake, in prior work with Shiu, Tang, and Torres, introduced the \emph{mixed volume of a chemical reaction network} 
    and Gross and Hill introduced a related~\emph{steady-state degree}. 
    We refer the reader to~\cite[Chapter~6]{NO-dissertation} for a discussion of the several ``flavors'' of mixed volume theory applied to chemical reaction networks. 
    These 
    parameter-free numerical invariants gave good bounds on the maximum number of steady states for several 
    families of networks~\cite{OSTT,mv-small-networks,gross-hill}. 
    However, the mixed-volume bound on a reaction network (and the related steady-state degree) is not always tight: the maximum number of \emph{observable, positive} steady states can be far less than the mixed volume. 
    Here, through our examples, we show that the self-intersection index can give a better, tighter bound than was previously possible. 
    
    In this paper, we explain the construction of a Newton-Okounkov body and the self-intersection index of a polynomial vector space, with an eye towards applying the theory to chemical reaction networks. 
    In an effort to make the theory of Newton-Okounkov bodies more accessible to an applied audience, we explain background material through a motivating example (the \textit{Wnt network}, see Example~\ref{ex:motivating-crn-sec}). 
    Moreover, our main contribution is one of the first concrete applications of Newton-Okounkov body theory to polynomial-system solving. 
    Readers interested in computing an upper bound on the number of effective solutions to a (parametrized) polynomial system may proceed to Procedure~\ref{proc:nob-compute}. 
    We emphasize that this procedure is not just for chemical reaction networks, and can be applied to any sparse parametrized polynomial system.

    The outline of this paper is as follows. 
    Section~\ref{sec:CRS} covers the basics of chemical reaction networks. 
    Section~\ref{sec:nob-setup} explains the Newton-Okounkov body theory and self-intersection index results we will need for our application. 
    Section~\ref{sec:nob-def-crn} explains our procedure for explicitly computing 
    a 
    self-intersection index of a chemical reaction network. 
    In Section~\ref{sec:nob-examples}, we 
    apply our procedure to several 
    chemical reaction networks. 
    We compare the resulting ``Newton-Okounkov body bound'' on the maximum number of steady states with both the mixed-volume bound and the actual maximum number of positive real steady states. 
    The paper concludes with a discussion of the implications of this new tool for chemical reaction network theory and other applications. 
    We include some conjectures and avenues for future research. 
    Finally, Appendix~\ref{app:notation} collects the notation introduced in Section~\ref{sec:nob-setup}.

    Supporting computations are archived in the following GitHub repository:~\url{https://github.com/EliseAWalker/NOBodies-for-ChemicalRxns}

\section{Chemical reaction networks and their dynamical systems} \label{sec:CRS}
    
    Our notation for chemical reaction networks closely matches that of%
    ~\cite{CFMW,DPST}. 
    We briefly review the basics of the mathematical models of chemical reaction networks in this section, and refer the reader to~\cite{feinberg,dickenstein-invitation} and references therein for background.

    A \defword{(chemical) reaction network} $G$  (or {\em network} for short) comprises a set of $d$ species $\{X_1, X_2, \dots, X_d\}$ and a set of $m$ reactions:
    \[
    \alpha_{1j}X_1 + 
    \alpha_{2j}X_2 +  \dots +
    \alpha_{dj}X_d
    ~ \overset{\kappa_j}{\longrightarrow} ~ %
    \beta_{1j}X_1 + 
    \beta_{2j}X_2 +  \dots +
    \beta_{dj}X_d~,
     \quad \quad
        {\rm for}~
    	j=1,2, \dots, m~,
    \]
    where each $\alpha_{ij}$ and $\beta_{ij}$ is a non-negative integer
    called a \defword{stoichiometric coefficient} and each $\kappa_j$ is a nonnegative real number called a \defword{reaction rate constant}. 
    The \defword{stoichiometric matrix} of 
    $G$, 
    denoted by $N$, is the $d\times m$ matrix with
    $(i, j)$-entry equal to $\beta_{ij}-\alpha_{ij}$. 
    Let $p=d-{\rm rank}(N)$. 
    The \defword{stoichiometric subspace}, denoted by $S$,
    is the image of $N$, that is, $S$ is the vector subspace of $\R^p$ generated by 
    the columns of $N$. 
    A \defword{conservation-law matrix} of $G$, denoted by $W$, is a row-reduced $(p\times d$)-matrix whose rows form a basis of 
    the orthogonal complement of $S$. 
    If there exists a choice of $W$ for which every entry is nonnegative and each column 
    contains at least one nonzero entry (equivalently, if 
    each species occurs in at least one nonnegative conservation law), then $G$ is \defword{conservative}.

    We denote the concentrations of the species $X_1,X_2, \dots, X_d$ by $x_1, x_2, \dots, x_d$, respectively. 
    These concentrations, under the assumption of {\em mass-action kinetics}, evolve in time according to the following system of 
    ordinary differential equations: 
    \begin{equation}\label{eq:ode}
    \dot{x}~=~f(x)~:=~N\cdot \begin{pmatrix}
    \kappa_1 \, x_1^{\alpha_{11}} 
    		x_2^{\alpha_{21}} 
    		\cdots x_d^{\alpha_{d1}} \\
    \kappa_2 \, x_1^{\alpha_{12}} 
    		x_2^{\alpha_{22}} 
    		\cdots x_d^{\alpha_{d2}} \\
    		\vdots \\
    \kappa_m \, x_1^{\alpha_{1m}} 
    		x_2^{\alpha_{2m}} 
    		\cdots x_d^{\alpha_{dm}} \\
    \end{pmatrix}~,
    \end{equation}
    where $x=(x_1, x_2, \dots, x_d)$, 
    and each $\kappa_j \in \mathbb R_{>0}$.
     By considering the rate constants as a vector of parameters $\kappa=(\kappa_1, \kappa_2, \dots, \kappa_m)$, we have polynomials $f_{\kappa,i} \in \mathbb Q[\kappa,x]$, for $i=1,2, \dots, d$.
    For ease of notation, we often write $f_i$ rather than  $f_{\kappa,i}$. 
    We refer to the dynamical system~\eqref{eq:ode} obtained from a chemical reaction network governed by mass-action kinetics as a \defword{chemical reaction system}. %

    A trajectory $x(t)$ beginning at a 
        positive vector $x(0)=x^0 \in
        \mathbb{R}^d_{> 0}$ 
    remains, for all positive time,
     in the following \defword{stoichiometric compatibility class} with respect to the \defword{total-constant vector} $c\coloneqq W x^0 \in {\mathbb R}^p$: %
    \begin{align} \label{eqn:invtPoly}
    \scc_c~\coloneqq~ \{x\in {\mathbb R}_{\geq 0}^d \mid Wx=c\}~.
    \end{align}
    Note that when $d={\rm rank}(N)$, the network has no conservation laws. In this case, the stoichiometric subspace $S$ is $\R^d$, and the only stoichiometric compatibility class~\eqref{eqn:invtPoly} is $\R_{\geq 0}^d$. 

\subsection{Steady states}\label{sec:def-steady-states}
    
    As summarized in the introduction, 
    one goal of 
    chemical reaction network theory is 
    to understand the fixed points, 
    called \emph{steady states}, 
    of %
    the differential equations~\eqref{eq:ode} defining chemical reaction systems.%
    
    A \defword{steady state} of~\eqref{eq:ode} is a 
    nonnegative 
    concentration vector 
    $x^* \in \mathbb{R}_{\geq 0}^d$ at which the 
    right-hand sides of the 
    chemical reaction system~\eqref{eq:ode}
    vanish: $f(x^*) =0$.  
    We distinguish between \defword{positive steady states} $x ^* \in \mathbb{R}^d_{> 0}$ and \defword{boundary steady states} 
    $x^*\in {\mathbb R}_{\geq 0}^d\backslash {\mathbb R}_{>0}^d$. 
    In particular, a boundary steady state $x^*\not\in (\C^*)^d$.  
    A network exhibits \defword{multistationarity} 
    if there exists a positive rate-constant vector $\kappa \in \mathbb{R}^m_{>0}$ such that there exist two or more positive steady states of~\eqref{eq:ode} in some stoichiometric compatibility class~\eqref{eqn:invtPoly}.  
    
    To analyze steady states within a stoichiometric compatibility class, we will 
    use conservation laws 
    in place of linearly dependent steady-state equations, as follows.
    Let $I = \{i_1 < i_2< \dots < i_p\}$ denote the indices of the first nonzero coordinate of the rows of conservation-law matrix $W$.
    Consider the function $\F: {\mathbb R}_{\geq 0}^d\rightarrow {\mathbb R}^d$ defined by 
    \begin{equation}\label{consys}
    f_{c,\kappa,i} =\F(x)_i :=
    \begin{cases}
    f_{i}(x)&~\text{if}~i\not\in I,\\
    (Wx-c)_k &~\text{if}~i~=~i_k\in I .
    \end{cases}
    \end{equation}
    We call system~\eqref{consys}, \defword{the system augmented by conservation laws} or just the \defword{augmented system}. By construction, positive roots of the system of polynomial equations $\F=0$ are precisely the positive steady states of~\eqref{eq:ode} in the stoichiometric compatibility class~\eqref{eqn:invtPoly} defined by the total-constant vector $c$.
    When the network has no conservation laws, then the augmented system is just the chemical reaction system~\eqref{eq:ode}.
    
    We conclude this section with an example of a chemical reaction network, the \emph{Wnt network} (Figure~\ref{fig:wnt-network}), which will serve as the motivating example throughout this paper. 

    \begin{example}[Motivating Example]\label{ex:motivating-crn-sec}
    The \defword{Wnt network} was introduced in~\cite{MacLean-Wnt} as a shuttle model of the Wnt signalling pathway, and it was systematically studied using algebraic-geometric methods in~\cite{wnt}.
    Depicted in Figure~\ref{fig:wnt-network}, the network comprises 19 chemical species (whose concentrations are denoted by the 19 variables $x_1,\dots,x_{19}$) participating in 31 reactions, labeled by $k_1,\dots,k_{31}$. 
    The network is not conservative (the network contains inflow and outflow reactions), however, the network has 5 conserved quantities $c_1,\dots,c_5$, and 
    subsequently 
    it has 
    5 linear conservation laws. 
    \begin{figure}[!ht]
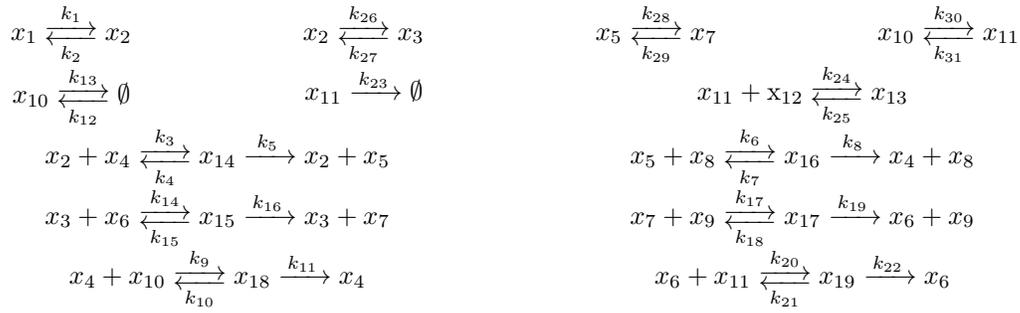

        \renewcommand{\arraystretch}{1.4}
        \begin{tabular}{M{0.21\linewidth} M{0.21\linewidth} M{0.21\linewidth} M{0.21\linewidth}}
        \centering
        $\ce{$x_1$ <-->[$k_1$][$k_2$] $x_2$}$     &          $\ce{$x_2$ <-->[$k_{26}$][$k_{27}$] $x_3$}$ &          $\ce{$x_5$ <-->[$k_{28}$][$k_{29}$] $x_7$}$ &          $\ce{$x_{10}$ <-->[$k_{30}$][$k_{31}$] $x_{11}$}$\\
        $\ce{$x_{10}$ <-->[$k_{13}$][$k_{12}$] \emptyset}$ & $\ce{$x_{11}$ ->[$k_{23}$] \emptyset}$ & \multicolumn{2}{c}{$\ce{$x_{11}$ + x_{12} <-->[$k_{24}$][$k_{25}$] $x_{13}$}$}\\    
        \multicolumn{2}{c}{$\ce{$x_2$ + $x_4$ <-->[$k_3$][$k_4$] $x_{14}$ ->[$k_5$] $x_2$ + $x_5$}$} &  \multicolumn{2}{c}{$\ce{$x_5$ + $x_8$ <-->[$k_6$][$k_7$] $x_{16}$ ->[$k_8$] $x_4$ + $x_8$}$}  \\
        \multicolumn{2}{c}{$\ce{$x_3$ + $x_6$ <-->[$k_{14}$][$k_{15}$] $x_{15}$ ->[$k_{16}$] $x_3$ + $x_7$}$} &  \multicolumn{2}{c}{$\ce{$x_7$ + $x_9$ <-->[$k_{17}$][$k_{18}$] $x_{17}$ ->[$k_{19}$] $x_6$ + $x_9$}$}  \\        
        \multicolumn{2}{c}{$\ce{$x_4$ + $x_{10}$ <-->[$k_9$][$k_{10}$] $x_{18}$ ->[$k_{11}$] $x_4$}$} &  \multicolumn{2}{c}{$\ce{$x_6$ + $x_{11}$ <-->[$k_{20}$][$k_{21}$] $x_{19}$ ->[$k_{22}$] $x_6$}$}  \\ %
        \end{tabular}
        \caption{The Wnt network, a formalism of the shuttle model of the Wnt signalling pathway~\cite{MacLean-Wnt,wnt}.}%
        \label{fig:wnt-network}
    \end{figure}
    
    Let $x(t) = (x_1(t),x_2(t),\dots,x_d(t))\in \Rnn^d$
    denote the vector of concentrations of the species as a differentiable function of time $t$. 
    Let $\kappa:=(k_1,\dots,k_{31})$ denote the vector of rate constants, and $c:=(c_1,\dots,c_5)$ denote the vector of conserved quantities. 
    In the chemical reaction system~\eqref{eq:ode} of the Wnt network arising from mass-action kinetics, the differential equations $\dot{x}_1,\dot{x}_4,\dot{x}_8,\dot{x}_9,\dot{x}_{12}$ are linear combinations of the remaining $\dot{x}_i$.
    This implies the 5 linear conservation equations $f_{c,\kappa}(x)_{1},~f_{c,\kappa}(x)_{4},~f_{c,\kappa}(x)_{8},~f_{c,\kappa}(x)_{9},~f_{c,\kappa}(x)_{12}$ seen in~\eqref{eq:wnt-aug-sys}. 
    The augmented system~\eqref{consys} of the Wnt network arising from mass-action kinetics is:%
    \begin{equation}\label{eq:wnt-aug-sys}
    \resizebox{.78\textwidth}{!}{$%
        \begin{cases}
            f_{1}:=f_{c,\kappa}(x)_1=x_1+x_2+x_3+x_{14}+x_{15}-c_1\\
            f_{2}:=f_{c,\kappa}(x)_2=-k_3 x_2 x_4+k_1 x_1+(-k_2-k_{26}) x_2+k_{27} x_3+(k_4+k_5) x_{14}\\
            f_{3}:=f_{c,\kappa}(x)_3=-k_{14} x_3 x_6+k_{26} x_2-k_{27} x_3+(k_{15}+k_{16}) x_{15}\\
            f_{4}:=f_{c,\kappa}(x)_4=x_4+x_5+x_6+x_7+x_{14}+x_{15}+x_{16}+x_{17}+x_{18}+x_{19}-c_2\\
            f_{5}:=f_{c,\kappa}(x)_5=-k_6 x_5 x_8-k_{28} x_5+k_{29} x_7+k_5 x_{14}+k_7 x_{16}\\
            f_{6}:=f_{c,\kappa}(x)_6=-k_{14} x_3 x_6-k_{20} x_6 x_{11}+k_{15} x_{15}+k_{19} x_{17}+(k_{21}+k_{22}) x_{19}\\
            f_{7}:=f_{c,\kappa}(x)_7=-k_{17} x_7 x_9+k_{28} x_5-k_{29} x_7+k_{16} x_{15}+k_{18} x_{17}\\  
            f_{8}:=f_{c,\kappa}(x)_8=x_8+x_{16}-c_3\\
            f_{9}:=f_{c,\kappa}(x)_9=x_9+x_{17}-c_4\\       
            f_{10}:=f_{c,\kappa}(x)_{10}=-k_9 x_4 x_{10}+(-k_{13}-k_{30}) x_{10}+k_{31} x_{11}+k_{10} x_{18}+k_{12}\\
            f_{11}:=f_{c,\kappa}(x)_{11}=-k_{20} x_6 x_{11}-k_{24} x_{11} x_{12}+k_{30} x_{10}+(-k_{23}-k_{31}) x_{11}+k_{25} x_{13}+k_{21} x_{19}\\
            f_{12}:=f_{c,\kappa}(x)_{12}=x_{12}+x_{13}-c_5\\            
            f_{13}:=f_{c,\kappa}(x)_{13}=k_{24} x_{11} x_{12}-k_{25} x_{13}\\
            f_{14}:=f_{c,\kappa}(x)_{14}=k_3 x_2 x_4+(-k_4-k_5) x_{14}\\
            f_{15}:=f_{c,\kappa}(x)_{15}=k_{14} x_3 x_6+(-k_{15}-k_{16}) x_{15}\\
            f_{16}:=f_{c,\kappa}(x)_{16}=k_6 x_5 x_8+(-k_7-k_8) x_{16}\\
            f_{17}:=f_{c,\kappa}(x)_{17}=k_{17} x_7 x_9+(-k_{18}-k_{19}) x_{17}\\
            f_{18}:=f_{c,\kappa}(x)_{18}=k_9 x_4 x_{10}+(-k_{10}-k_{11}) x_{18}\\
            f_{19}:=f_{c,\kappa}(x)_{19}=k_{20} x_6 x_{11}+(-k_{21}-k_{22}) x_{19}\\
        \end{cases}$%
        }
    \end{equation}

    \end{example}

\subsection{Bounds on the number of steady states}\label{sec:classical-bounds}

    As is true for many polynomial systems arising from applications, 
    chemical reaction systems are structured and sparse. 
    Consequently, the study of chemical reaction networks benefits from the study of sparse polynomial systems. 
    We conclude this section by recalling a few fundamental upper bounds on the number of isolated solutions to sparse polynomial systems, and their applications to reaction networks. 
    
    For 
    a polynomial $f = a_1 x^{\sigma_1} + a_2 x^{\sigma_2} + \dots + a_{\ell} x^{\sigma_{\ell}} \in \mathbb{R}[x_1,x_2,\dots, x_d]$, where 
    the 
    exponent vectors $\sigma_i \in \mathbb{Z}^d$ are distinct and 
    $a_i \neq 0$ for all $i$, %
    the \defword{Newton polytope} of $f$
    is the convex hull of its exponent vectors:
    ${\rm Newt}(f) \coloneqq {\rm conv} \{\sigma_1, \sigma_2,\dots ,\sigma_{\ell}\} \subseteq \mathbb{R}^d.$
    The (finite) set of exponent vectors $\{\sigma_1,\dots,\sigma_{\ell}\}$ is called the \defword{support} of $f$. 
    The \defword{Kushnirenko Theorem} gives 
    one classical bound on the maximum number of nonzero isolated complex solutions to a polynomial system where each polynomial has the same Newton polytope:
    
    \begin{prop}[Kushnirenko Theorem~\cite{kushnirenko-bound}]\label{prop:kushnirenko}
    Consider $d$ real polynomials $g_1, g_2, \dots, g_d \in \mathbb{R}[x_1, x_2, \dots, x_d]$ with common support, and let $G:=\newt{g_i}$ for all $i$. 
    Then the number of isolated solutions in $(\C^*)^d$, counted with multiplicity, of the system 
    $g_1(x) = g_2(x) = \cdots = g_d(x) = 0$ is at most $d!~\vol{G}$.         
    \end{prop}
    
    In order to state a generalization of this result, we will need the following definition. 
    
    \begin{defn}\label{def:mink-sum} 
    Let $Q_1, Q_2, \dots,Q_d\subseteq \R^d$ be polytopes. 
    The normalized volume of the Minkowski sum $\lambda_1Q_1+ \lambda_2 Q_2+ \dots+\lambda_d Q_d$ is a homogeneous polynomial of degree $d$ in nonnegative variables $\lambda_1,\lambda_2,\dots,\lambda_d$. In this polynomial, the coefficient 
    of $\lambda_1 \lambda_2\cdots\lambda_d$, 
    denoted by \defword{$\mv{Q_1, Q_2, \dots,Q_d}$}, is the \defword{mixed volume} of $Q_1,Q_2,...,Q_d$. 
    \end{defn}

    The \defword{Bernstein}-Khovanskii-Kushnirenko \defword{theorem} uses the mixed volume of the Newton polytopes of a polynomial system to count its solutions in $(\C^*)^d$.%
    
    \begin{prop}[Bernstein Theorem~\cite{bernstein}] \label{prop:bernstein} 
    Consider $d$ real polynomials $g_1, g_2, \dots, g_d \in \mathbb{R}[x_1, x_2, \dots, x_d] $. 
    Then the number of isolated solutions in $(\C^*)^d$, counted with multiplicity, of the system 
    $g_1(x) = g_2(x) = \cdots = g_d(x) = 0$ is at most $\mv{\newt{g_1}, \dots , \newt{g_d}}$. 
    \end{prop}

    The Kushnirenko Theorem and its generalization, the 
    Bernstein Theorem, connect convex geometry with algebraic geometry: the volumes of polytopes are used to count the number of solutions in $(\C^*)^d$ to a generic polynomial system 
    drawn from the 
    polynomial vector space spanned by the monomials with exponent vectors in the relevant polytopes. 
    Proofs of these bounds leverage rich theory from algebraic geometry. 
    Recently, Kaveh and Khovanskii~\cite{kk-nob-1} realized the Kushnirenko and Bernstein Theorems as special cases of their more general result, as we will see in Section~\ref{sec:kk-thm}. 

    Motivated by Proposition~\ref{prop:bernstein}, \cite{OSTT} introduced the definition of the mixed volume of a chemical reaction network.

    \begin{defn}\label{def:mv-crn}
    Let $G$ be a network with $d$ species, 
    $m$ reactions, and 
    a $p \times d$ conservation-law matrix $W$, 
    which results in the system augmented by conservation laws 
    $f_{c,\kappa}$, as in~\eqref{consys}.  
    Let $c^*\in \mathbb{R}^p_{\neq 0}$, and let 
    $\kappa^* \in \mathbb{R}^m_{>0}$ be generic. 
    Let $Q_1, Q_2, \dots,Q_d\subset \R^d$ be the Newton polytopes of $f_{c^*,\kappa^*,1},f_{c^*,\kappa^*,2}, \dots, f_{c^*,\kappa^*,d}$, respectively. The \defword{mixed volume of $G$} (with respect to $W$)
    is the mixed volume of $Q_1,Q_2,\dots,Q_d$.  
    \end{defn}

    Every positive steady state is a steady state over $\mathbb{C}^*$.
    Also, the mixed volume pertains to polynomial systems with the same supports 
    as the augmented system $f_{c,\kappa}=0$ (but without constraining the coefficients to come from a reaction network).  We obtain, therefore, the bounds in the following result:
    
    \begin{prop}[\cite{OSTT}]%
    \label{prop:bounds}
    For every reaction network, the following inequalities hold among the
    maximum number of positive steady states, 
    the maximum number of steady states over $\mathbb{C}^*$, and 
    the mixed volume of the network (with respect to any conservation-law matrix):
    \[
    \mathrm{max~\#~of~positive~steady~states}
    ~\leq ~ 
    \mathrm{max~\#~of~steady~states~over~} \mathbb{C}^*
    ~\leq ~
    {\rm mixed~volume}~.
    \]
    \end{prop}

    Note that some networks have \textit{boundary} steady states, 
    but boundary steady states are \textit{not} counted by mixed volume.
    
    As an example, we now compute the mixed volume of our motivating example (also computed in~\cite{NO-dissertation}). 
    (Other related mixed volumes were also computed in~\cite{wnt}.)
    
    \begin{example}[Motivating Example, continued]\label{ex:motivating-bkk-sec}
    Recall the Wnt network from Example~\ref{ex:motivating-crn-sec}. 
    Let $e_i$ denote the $i$-th standard basis vector of $\R^{19}$. 
    The Newton polytope of $f_1$ is $\conv\{e_1,e_2,e_{14},e_{15}\}$, the Newton polytope of $f_2$ is $\conv\{e_2+e_4,e_1,e_2,e_3,e_{14}\}$, and 
    we can compute the remaining Newton polytopes $\newt{f_3},\dots,\newt{f_{19}}$ analogously. 
    By Definition~\ref{def:mv-crn}, the mixed volume of $G$ is 56, and by Proposition~\ref{prop:bounds}, we conclude that $G$ has at most 56 positive steady states (for a generic choice of $\kappa,c$). 
    Supporting computations are archived in the repository~\url{https://github.com/EliseAWalker/NOBodies-for-ChemicalRxns}
    \end{example}

    \begin{table}[ht!]
        \centering
        \begin{tabular}{p{0.2\linewidth}|p{0.18\linewidth} p{0.18\linewidth} p{0.25\linewidth}}
            network & max number of positive steady states & mixed volume & max number of solutions over $\C^*$ \\ \midrule
            fully irreversible {ERK} & 1~\cite{OSTT} & 3 & 3 (attained over $\R^*$~\cite{OSTT})\\
            full {ERK} & 3 (conjecture)~\cite{OSTT} & 7 & 7 (attained over $\C^*$~\cite{OSTT})\\
            Edelstein &3 & 3 &3~\cite{gross-hill}\\
            Wnt & 3 (conjecture)~\cite{wnt} & 56~(Example~\ref{ex:motivating-bkk-sec}) & 9 (attained over $\R$~\cite{wnt})

        \end{tabular}
        \caption{Selected results on mixed volume of chemical reaction networks and related bounds on the maximum number of positive steady states.}
        \label{tab:mv-crn-past-results}
    \end{table}

    Table~\ref{tab:mv-crn-past-results} catalogs the maximum number of steady states, the mixed volume, and the computed maximum number of nonzero complex steady states for selected chemical reaction networks.     
    For some networks, like the fully irreversible {ERK} network~\cite{OSTT} and the Edelstein network~\cite{gross-hill,apoptosis-edelstein-mfplv}, the mixed volume gives a good bound on the maximum number of observable positive steady states. 
    For others, like the {Wnt} network, the mixed volume greatly exceeds the maximum number of observable steady states, motivating the need for a much sharper bound. 
    Indeed, 
    using Gr\"obner-basis methods, 
    \cite{wnt} showed that the {Wnt} model has a maximum number of 9 complex steady states. 
    Moreover,~\cite{wnt} numerically provided a witness of 3 \emph{positive} steady states, and thus conjectured that the maximum number of positive steady states cannot exceed 3. 
    We conclude that the mixed-volume bound overcounts the actual number of observable steady states by at least $45$. 
    Can we do better? 
    In this paper, we introduce a complementary approach to those in~\cite{wnt}, adding a new algebraic-geometric tool for studying chemical reaction networks.  
    In the next section, we will show that for the {Wnt} network, the resulting bound is tighter than the mixed-volume bound.

\section{Newton-Okounkov body bound background}\label{sec:nob-setup}
    
In~\cite{kk-nob-1,kaveh-khovanskii-first-def-biii-2010}, 
Kaveh and Khovanskii investigated a generalization of the mixed-volume bound. That is, instead of using the monomial-support system to give a bound on the number of solutions, they instead considered systems with \emph{polynomial} support. The bound they gave is the \emph{birationally invariant intersection index}. They furthermore gave a formula for the \emph{self-intersection index} using Newton-Okounkov bodies. We will consequently refer to this bound as the \emph{Newton-Okounkov body bound}. This bound is complicated to define and compute. This section concerns itself with the necessary definitions and concepts.
        
    We begin with a brief overview of 
    the setup. 
    Let $X$ be a $d$-dimensional algebraic variety and let $V\subset \C(X)$ be a $n$-dimensional vector space with finite \emph{Khovanskii basis} $\mathcal{B}=\{b_0,\dots,b_n\}$. 
    The rational \emph{Kodaira map} $\phi_V:X\dashrightarrow\mathbb{P}(V')\simeq \mathbb{P}^n$, defined by $x\mapsto [v_0(x),\dots,v_n(x)]$ parametrizes a variety as follows. 
    The Zariski closure of the image of $\phi_V$ is a projective variety; call it $\kodproj:=\overline{\phi_V(X)}$. 
    The coordinate ring of $\kodproj$ is 
    a graded ring 
    $\rv$. 
    Given a finite Khovanskii basis for $\rv$, one may compute a \emph{Newton-Okounkov body} of $\rv$. 
    Convex bodies -- Newton polytopes and Newton-Okounkov bodies -- associated to projective varieties give us a beautiful way to compute %
    the \emph{self-intersection index} of $V$. 
    Next, we define the objects needed to construct a Newton-Okounkov body associated to 
    a vector space $V$ of functions defined on a projective variety.
  
    \subsection{Polynomial vector space, graded algebra}\label{sec:vs-graded-alg}
    A (nonempty, finite-dimensional) subspace $V\subset \C(X)$ gives rise to a graded algebra $\rv$ as follows. 
    Let $s$ be a formal grading variable, and let $V^k \subset \C(X)$ denote the subspace generated by all $k$-fold products of elements in $V$, for $k\in \N$ (here $V^0$ is just the base field $\C$). 
    Any nonzero element $f$ in the $\C$-algebra generated by $Vs$  
    can be decomposed into a sum of $s$-homogeneous components: $f=f_ks^k+\dots+f_1s+f_0$, where $f_k\neq 0$ and $f_j\in V^j$ for each $j$. 
    Decomposing $f\in\rv$ in this way, we see that $\rv \subset \C(X)[s]$, and $\rv$ is graded by the highest degree of $s$. 
    Let \defword{$\rv$}$:=\bigoplus_{k\geq 0} V^ks^k$ denote this \defword{graded algebra associated to $V$}. 
    
    \begin{example}[Motivating Example, continued]\label{ex:motivating-example-vect-space-graded-alg}
        In this example, we associate a finite-dimensional vector space $V$ to the {Wnt} network. 
        We also compute a finite Khovanskii basis of the corresponding algebra $\rv$. 
        Consider the following vector subspace $V$ of $\C(X)$ spanned by 29 generators $v_0,\dots,v_{28}$: 
        \begin{align*}
            V = \spn_{\C}\{
              &-k_3 x_2 x_4+ (-k_2-k_{26}) x_2+ (k_4+k_5) x_{14},~
              k_1 x_1+ k_{27} x_3,~
              -k_{14} x_3 x_6 -k_{27} x_3 +( k_{15}+ k_{16}) x_{15},\\
              &k_{26} x_2,~
              -k_6 x_5 x_8 -k_{28} x_5+k_7 x_{16},~
              k_{29} x_7+k_5 x_{14},~
              -k_{14} x_3 x_6 -k_{20} x_6 x_{11}+ k_{15} x_{15},~\\
               &k_{19} x_{17}+( k_{21}+ k_{22}) x_{19},~
               -k_{17} x_7 x_9-k_{29} x_7+ k_{18} x_{17},\\
              &k_{28} x_5+ k_{16} x_{15},~
              -k_9 x_4 x_{10}+(-k_{13}-k_{30}) x_{10}+ k_{10} x_{18},~
              k_{31} x_{11}+ k_{12},\\
              &-k_{24} x_{11} x_{12} + (-k_{23}-k_{31}) x_{11} + k_{25} x_{13},~
              -k_{20} x_6 x_{11}+ k_{21} x_{19}+k_{30} x_{10},~
               k_{24} x_{11} x_{12}-k_{25} x_{13},~\\
              &k_3 x_2 x_4+(-k_4-k_5) x_{14},~
               k_{14} x_3 x_6+(-k_{15}-k_{16}) x_{15},~
              k_6 x_5 x_8+(-k_7-k_8) x_{16},\\
               &k_{17} x_7 x_9+(-k_{18}-k_{19}) x_{17},~
              k_9 x_4 x_{10}+(-k_{10}-k_{11}) x_{18},~
               k_{20} x_6 x_{11}+(-k_{21}-k_{22}) x_{19},~\\
              &x_2+x_{14},~
              x_3+x_{15},~
              x_1-c_1,~
              x_4+x_6+x_{15}+x_{19}+x_{14}+x_{18},~
              x_5+x_{16}+x_7+x_{17}-c_2,~\\
              &x_8+x_{16}-c_3,~
              x_9+x_{17}-c_4,~
              x_{12}+x_{13}-c_5~
                    \}
                    =:~
                    \spn_{\C}\{v_0,\dots,v_{28}\}
                    ~.
    \end{align*}

    \noindent Note that $V$ is chosen such that the defining polynomials $f_i$ of the Wnt network's augmented system~\eqref{eq:wnt-aug-sys} are $\C$-linear combinations of the basis polynomials $v_j$. 
    Let $\rv=\bigoplus_{k\geq 0} V^k s^k$ be the graded 
    algebra associated to $V$. 
    \end{example}

    \subsection{Degree of Kodaira map}\label{sec:deg-kodaira-map}
    \begin{definition}[Kodaira map]\label{def:kodaira-map-using-vs-basis}
    
    Given a vector space $W$ over a field $F$ its dual space $W'$ is the $F$-vector space 
    of all linear functionals $W\rightarrow F$. 
    The \defword{projectivization of an $F$-vector space $W$} is \defword{$\pp(W)$} $:= W'/\sim$ where $\sim$ is the equivalence relation $u \sim v$ if $u=\lambda v$ for $u,v\in W'$ and $\lambda\in F^*$. 
    
        Let $V \subset \C(X)$ be nonzero finite dimensional vector subspace. 
        Suppose $\{v_0,\ldots,v_n\}$ is a vector space 
        basis for $V$. 
        Then for $x\in X$ such that $v_i(x)$ is defined for all $i$ and $v_i(x)\neq 0$ for some $i$, 
        the rational \defword{Kodaira map $\varphi_V:X\dashrightarrow \pp(V')$ of the subspace $V$} is given by:
        \[\varphi_V(x)=[v_0(x):\dots:v_n(x)]~.\]    
    \noindent The \defword{degree $\text{deg}~\varphi_V$ of the Kodaira map $\varphi_V$} 
    is the maximum 
    number of isolated points in the inverse image of any point:
    $|\varphi_V^{-1}([z_0:\dots:z_n])|$. 
    The homogeneous coordinate ring of 
    the projective variety $\overline{\phi_V(X)}$ 
    is isomorphic to the graded algebra $\rv$~\cite{kk-nob-1}. %
    \end{definition}

    \begin{example}[Motivating Example, continued]\label{ex:motivating-Kodaira-map}
    Recall that a vector space basis for $V$ given in Example~\ref{ex:motivating-example-vect-space-graded-alg} is $\{v_0,\dots,v_{28}\}$. 
    The Kodaira map $\phi_V$ is defined by $x\mapsto [v_0(x): \dots: v_{28}(x)]$. 
    Using elimination techniques, we compute the degree $\deg(\phi_V)$ of the Kodaira map to be $1$.  
    \end{example}

    \subsection{Value semigroup}\label{sec:valuation}
    A \emph{valuation} is a map that translates subalgebra computations into computations in an associated semigroup. 
    We give the following definition for a valuation on a $\C$-algebra, $\C(X)$.
    
    \begin{definition}[valuation]\label{def:valuation}
        Let $\succ$ be a total order on $\Z^d$ that restricts to a well-ordering on $\Z^d_{\geq 0}$. 
        A \defword{$\Z^d$-valuation on $\C(X)$} is a 
        surjective group homomorphism 
        $\nu:\C(X) \rightarrow \Z^d \cup \{\infty\}$ 
        that satisfies the following properties for all $f,g\in \C(X)^*$ and $c\in\C^*$: 
        \begin{enumerate}
            \item $\nu(0)=\infty$, where $\infty \succ \alpha$, and $\alpha+\infty=\infty$ for all $\alpha \in \Z^d$, 
            \item $\nu(c)=(0,\dots,0)$, and  %
            \item $\nu(f+g)\succeq \text{min}\{\nu(f),\nu(g)\}$.
        \end{enumerate}

    \end{definition}
    
    Throughout this paper, we will restrict ourselves to a specific type of valuation induced by a monomial term order called a \emph{Gr\"obner valuation}. 
    \begin{defn}\label{def:grobner-valuation}
    Fix a total ordering on $\Z^d$. 
    Define a \defword{Gr\"obner valuation} $\nu$ on $\C(X)$ as follows. 
    For $f\in \C[X]$, let $cx_1^{a_1}\dots x_d^{a_d}$ be the term in $f$ with the smallest exponent $(a_1,\dots,a_d)$ with respect to the total ordering on $\Z^d$. 
    Define $\nu(f)=(a_1,\dots,a_d)$. 
    Extend $\nu$ to $\C(X)$ by defining $\nu(f/g)=\nu(f)-\nu(g)$ for any $f,g\in \C[X]$ with $g\neq 0$. 
    A Gr\"obner valuation $\nu:\C(X)\rightarrow \Z^d$ is a 
    $\Z^d$-valuation with one-dimensional leaves on $\C(X)$~\cite{kk-nob-1}. 
    \end{defn}
    
    \begin{defn}[grevlex~\cite{cox-little-oshea}]
    The \defword{g}raded \defword{rev}erse \defword{lex}icographic term order (\defword{grevlex})
    on the set of monomials in a (multivariate) polynomial ring $k[{\bf x}]$ is defined by: $x^A > x^B$ if either degree($x^A$) $>$ degree($x^B$) or degree($x^A$) = degree($x^B$) and the last nonzero entry of the vector of integers $A-B$ is negative. 
    \end{defn}
    
    Every monomial term order, including grvlex, is extendable to a Gr\"{o}bner valuation where the \emph{smallest} exponent in the valuation (in the total order $\succ$ on $\Z^d$) corresponds to the exponent of the \emph{leading} term of the monomial term order. %

    We extend a $\Z^d$-valuation $\nu$ on $\C(X)$ to a \defword{$\Z^d\oplus \N$-valuation} 
    $\widehat{\nu}:\rv\rightarrow \Z^d\oplus \N$ \defword{on the graded algebra~$\rv$} as follows: 
    for $f\in \rv^*$ with degree $k$ as a polynomial in $\C(X)[s]$, 
    \[\widehat{\nu}(f) = \widehat{\nu}(f_ks^k+\dots+f_1s+f_0) = (\nu(f_k), k)~.\]
    The image of $\rv^*$ under $\widehat{\nu}$ is defined to be the \defword{value semigroup} $S$ 
    of $(V,\nu)$:
    that is, $S:= S(V,\nu) = \{\widehat{\nu}(f)~|~f\in \rv^*\}$. 
    We remark that although $S$ is more precisely a monoid ($0\in S$ since $\nu(c)=0$ for any $c\in \C^*$), we follow the convention to refer to $S$ as a semigroup.  
    The \emph{rank} of a valuation $\widehat{\nu}$ is the rank of the sublattice of $\Z^d\oplus \N$ generated by the value semigroup $S(V,\nu)$. 
    In all cases we consider, the valuation has full rank, that is, $\text{rank}(\widehat{\nu})=d+1=\dim(S)$.

    \subsection{Khovanskii basis}\label{sec:khovanskii-basis}
    A Khovanskii basis is a special generating set of an algebra, first formalized in~\cite{kaveh-manon}. 
            
    \begin{definition}[Khovanskii basis~\cite{kaveh-manon}]\label{def:khovanskii-basis}
    Let $\nu$ be a valuation. %
    A 
    subset $\mathcal{B}\subset \rv$ is a \defword{Khovanskii basis} for $(V,\nu)$ 
    if the set $\{\nu(b)~|~b\in\mathcal{B}\}$ generates $S(V,\nu)$ as a semigroup. 
    \end{definition}
    
    It follows from the definition that a Khovanskii basis $\mathcal{B}$ generates $\rv$ as a $\C$-algebra, and $\mathcal{B}\cap Vs$ is a vector space basis for $Vs$.

    \begin{remark}[SubalgebraBases.m2]\label{rmk:khovanskii-basis-finite-assumption}
    We will only work with 
    graded polynomial 
    algebras $\rv$
    that have a \textit{finite} Khovanskii 
    basis with respect to some Gr\"obner valuation induced by a monomial term order. 
    Under this assumption (namely, a finite Khovanskii basis exists), 
    \cite{kaveh-manon} gives an algorithm for computing a finite Khovanskii basis $\mathcal{B}=\{b_1,\dots,b_n\}$ from a finite set of $k$-algebra generators for $\rv$. 
    The recent Macaulay2 package {\tt SubalgebraBases.m2}~\cite{subalgebrabases-m2} implements~\cite{kaveh-manon}'s algorithm; 
    we use {\tt SubalgebraBases.m2} to compute Khovanskii bases in our examples (Section~\ref{sec:nob-examples}). 
    \end{remark}

    \begin{example}[Motivating Example, continued]\label{ex:motivating-grobner-valuation}    
    Let $\nu$ be the Gr\"obner valuation on $\C(X)$ induced by the graded reverse lexicographic (grevlex) monomial term order on $\C[x_1,\dots,x_{19}]$.  
    We extend the Gr\"obner valuation on $\C(X)$ to $\C(X)[s]$, so that $\widehat{\nu}(f) = (\nu(f), \deg_s(f))$.  
    The vector space basis $v_0,\dots,v_{28}$ induces a vector space basis $v_0s,\dots,v_{28}s$ for $Vs$. 
    Using the vector space basis $v_0s,\ldots,v_{28}s$ for $Vs$ as input to {\tt SubalgebraBases.m2}, we compute the following Khovanskii basis $\mathcal{B}$ (with respect to $\nu$) for $\rv$:%
    \begin{align}\label{eq:wnt-khovanskii-basis}
        \Big\{    &s,\,{x}_{19}s,\,{x}_{18}s,\,{x}_{17}s,\,{x}_{16}s,\,{x}_{15}s,\,{x}_{14}s,\,{x}_{12}s+{x}_{13}s,\,{x}_{11}s,\,{x}_{10}s,\,{x}_{9}s,\,{x}_{8}s,\,{x}_{7}s,\,{x}_{5}s,\,{x}_{4}s+{x}_{6}s,\,{x}_{3}s,\,
        \notag \\ 
    &{x}_{2}s,\,{x}_{1}s,\,{x}_{11}{x}_{12}s+\frac{-{k}_{25}}{{k}_{24}}\,{x}_{13}s, 
    \,{x}_{6}{x}_{11}s,\,{x}_{4}{x}_{10}s,\,{x}_{7}{x}_{9}s,\,{x}_{5}{x}_{8}s,\,{x}_{3}{x}_{6}s,\,{x}_{2}{x}_{4}s,\,{x}_{11}{x}_{13}s^{2}+\frac{{k}_{25}}{{k}_{24}}\,{x}_{13}s^{2},\, 
    \notag \\
    &{x}_{6}{x}_{10}s^{2},\,{x}_{2}{x}_{6}s^{2},\,{x}_{6}{x}_{10}{x}_{11}{x}_{13}s^{3}+\frac{{k}_{25}}{{k}_{24}}\,{x}_{6}{x}_{10}{x}_{13}s^{3},\,{x}_{2}{x}_{6}{x}_{11}{x}_{13}s^{3}+\frac{{k}_{25}}{{k}_{24}}\,{x}_{2}{x}_{6}{x}_{13}s^{3},\,
    \notag \\
    \begin{split}   
    {}&{x}_{2}{x}_{4}{x}_{6}{x}_{11}^{2}{x}_{13}^{2}s^{5}+{x}_{2}{x}_{6}^{2}{x}_{11}^{2}{x}_{13}^{2}s^{5}+\frac{2\,{k}_{25}}{{k}_{24}}\,{x}_{2}{x}_{4}{x}_{6}{x}_{11}{x}_{13}^{2}s^{5}\\
    &\qquad\qquad\qquad\qquad\qquad\qquad\qquad\qquad +\frac{2\,{k}_{25}}{{k}_{24}}\,{x}_{2}{x}_{6}^{2}{x}_{11}{x}_{13}^{2}s^{5}+\frac{{k}_{25}^{2}}{{k}_{24}^{2}}\,{x}_{2}{x}_{4}{x}_{6}{x}_{13}^{2}s^{5}+\frac{{k}_{25}^{2}}{{k}_{24}^{2}}\,{x}_{2}{x}_{6}^{2}{x}_{13}^{2}s^{5},\,\end{split}\notag \\  
    \begin{split}  &{x}_{4}{x}_{6}{x}_{10}{x}_{11}^{2}{x}_{13}^{2}s^{5}+{x}_{6}^{2}{x}_{10}{x}_{11}^{2}{x}_{13}^{2}s^{5}+\frac{2\,{k}_{25}}{{k}_{24}}\,{x}_{4}{x}_{6}{x}_{10}{x}_{11}{x}_{13}^{2}s^{5}\\
    &\qquad\qquad\qquad\qquad\qquad\qquad\qquad\qquad +\frac{2\,{k}_{25}}{{k}_{24}}\,{x}_{6}^{2}{x}_{10}{x}_{11}{x}_{13}^{2}s^{5}+\frac{{k}_{25}^{2}}{{k}_{24}^{2}}\,{x}_{4}{x}_{6}{x}_{10}{x}_{13}^{2}s^{5}+\frac{{k}_{25}^{2}}{{k}_{24}^{2}}\,{x}_{6}^{2}{x}_{10}{x}_{13}^{2}s^{5} \Big\}\end{split}
    \end{align}

    The Khovanskii basis has 32 generators, which we denote $b_0,\dots,b_{31}$. %
    For each polynomial, the leading monomial (with respect to grevlex) of each basis vector is the first monomial listed. 
    By definition, the semigroup $S(V,\nu)$ is generated by the valuations $\widehat{\nu}(b_i)$ of the Khovanskii basis elements. 
    Let $e_i$ denote the $i$-th standard basis vector for $\R^{19}$. 
    Then, for example, 
    $\widehat{\nu}\left({x}_{11}{x}_{12}s-({k}_{25}/{k}_{24
      })\,{x}_{13}s\right)=(e_{11}+e_{12},1)\in S(V,\nu)$ 
      since the Khovanskii basis element $b_{19} = {x}_{11}{x}_{12}s-({k}_{25}/{k}_{24
      })\,{x}_{13}s$ has leading term ${x}_{11}{x}_{12}s$ under grevlex. 
    The Gr\"obner valuation of the remaining $b_i$ are computed analogously. 
    The valuations of the 32 Khovanskii basis elements are the columns of the following matrix $A$: 
    \[
    \resizebox{.8\textwidth}{!}{
$\begin{pmatrix}
0&0&0&0&0&0&0&0&0&0&0&0&0&0&0&0&0&1&0&0&0&0&0&0&0&0&0&0&0&0&0&0\\
0&0&0&0&0&0&0&0&0&0&0&0&0&0&0&0&1&0&0&0&0&0&0&0&1&0&0&1&0&1&0&1\\
0&0&0&0&0&0&0&0&0&0&0&0&0&0&0&1&0&0&0&0&0&0&0&1&0&0&0&0&0&0&0&0\\
0&0&0&0&0&0&0&0&0&0&0&0&0&0&1&0&0&0&0&0&1&0&0&0&1&0&0&0&0&0&1&1\\
0&0&0&0&0&0&0&0&0&0&0&0&0&1&0&0&0&0&0&0&0&0&1&0&0&0&0&0&0&0&0&0\\
0&0&0&0&0&0&0&0&0&0&0&0&0&0&0&0&0&0&0&1&0&0&0&1&0&0&1&1&1&1&1&1\\
0&0&0&0&0&0&0&0&0&0&0&0&1&0&0&0&0&0&0&0&0&1&0&0&0&0&0&0&0&0&0&0\\
0&0&0&0&0&0&0&0&0&0&0&1&0&0&0&0&0&0&0&0&0&0&1&0&0&0&0&0&0&0&0&0\\
0&0&0&0&0&0&0&0&0&0&1&0&0&0&0&0&0&0&0&0&0&1&0&0&0&0&0&0&0&0&0&0\\
0&0&0&0&0&0&0&0&0&1&0&0&0&0&0&0&0&0&0&0&1&0&0&0&0&0&1&0&1&0&1&0\\
0&0&0&0&0&0&0&0&1&0&0&0&0&0&0&0&0&0&1&1&0&0&0&0&0&1&0&0&1&1&2&2\\
0&0&0&0&0&0&0&1&0&0&0&0&0&0&0&0&0&0&1&0&0&0&0&0&0&0&0&0&0&0&0&0\\
0&0&0&0&0&0&0&0&0&0&0&0&0&0&0&0&0&0&0&0&0&0&0&0&0&1&0&0&1&1&2&2\\
0&0&0&0&0&0&1&0&0&0&0&0&0&0&0&0&0&0&0&0&0&0&0&0&0&0&0&0&0&0&0&0\\
0&0&0&0&0&1&0&0&0&0&0&0&0&0&0&0&0&0&0&0&0&0&0&0&0&0&0&0&0&0&0&0\\
0&0&0&0&1&0&0&0&0&0&0&0&0&0&0&0&0&0&0&0&0&0&0&0&0&0&0&0&0&0&0&0\\
0&0&0&1&0&0&0&0&0&0&0&0&0&0&0&0&0&0&0&0&0&0&0&0&0&0&0&0&0&0&0&0\\
0&0&1&0&0&0&0&0&0&0&0&0&0&0&0&0&0&0&0&0&0&0&0&0&0&0&0&0&0&0&0&0\\
0&1&0&0&0&0&0&0&0&0&0&0&0&0&0&0&0&0&0&0&0&0&0&0&0&0&0&0&0&0&0&0\\
1&1&1&1&1&1&1&1&1&1&1&1&1&1&1&1&1&1&1&1&1&1&1&1&1&2&2&2&3&3&5&5\end{pmatrix}$
}
    \]
    
    By definition, since $\mathcal{B}$ is a Khovanskii basis for $(V,\nu)$, the semigroup $S(V,\nu)\subset \Z^{19}\bigoplus \N$ is generated by the columns of the matrix $A$.

    \end{example}

    \begin{remark}\label{finite-khovanskii-bassis-not-guaranteed}
    
    Deciding whether a finitely generated subalgebra has a finite Khovanskii basis is still an open question, in general. 
    We refer the interested reader to~\cite{robbiano1990subalgebra, sturmfels-algs-in-invariant-theory, kaveh-manon} for examples of 
    intricacies of computing Khovanskii bases. 
    \end{remark}

    \subsection{Index of graded algebra}\label{sec:ind-graded-algebra}
    In order to define the index associated to $\rv$, we will need a couple more definitions. 
    Let $G(S)\subset \R^{d+1}$ denote the group generated by the semigroup $S$, so 
    $G(S)~=~\left\{\sum_i k_i a_i~|~a_i\in S, k_i\in \Z\right\}$. 
    Let $\pi:\R^{d+1}\rightarrow \R$ denote projection onto the $(d+1)$-st coordinate. 
    Let $G_0(S)\subset G(S)$ denote the subgroup $\pi^{-1}(0)\cap G(S)$ of $\Z^d\oplus \{0\}$. 
    The \defword{index of $\rv$}, $\text{ind}(\rv)$, 
    is the index of the subgroup 
    $G_0(S)$ in $\Z^d \bigoplus \{0\}$. %

    \begin{example}\label{ex:pm-index-g0}
        Consider $\rv$ where $V=\spn\{x_1(x_1^2+x_2^2 -2x_1),~x_1(5-4x_2),~x_2(x_1^2+x_2^2 -2x_1),~x_2(5-4x_2)\}$. Using $s$ as a grading variable, this is equivalent to considering the algebra generated by $\{x_1(x_1^2+x_2^2 -2x_1)s,~x_1(5-4x_2)s,~x_2(x_1^2+x_2^2 -2x_1)s,~x_2(5-4x_2)s\}$ over $\C$. 
        
        Using the Grobner valuation induced by grevlex 
        as the valuation $\widehat{\nu}$, 
        one finds that the the algebra generators are also a finite Khovanskii basis for $\rv$ with respect to this valuation.
        Thus, the value semigroup $S$ is generated by the points: $(0,2,1), (1,1,1),(2,1,1), (3,0,1)$. 
        The group generated by $S$ is $\Z^2 \oplus \Z$. 
        By taking integer combinations of the generators  (specifically, pairwise sums of $\pm(0,2,1), \pm(1,1,1), \pm(2,1,1), \pm(3,0,1)$), then one finds that $G_0(R)$ contains the points $(1,-1,0), (-1,0,0), (1,0,0), \text{ and } (-1,1,0)$, which is enough to generate all of $\Z^2 \oplus \{0\}$. 
        Thus $G_0(R) = \Z^2 \oplus \{0\}$, and so $\ind(R) = [\Z^2 \oplus \{0\}: G_0(R)] = 1$.
    \end{example}

    \begin{example}[Motivating Example, continued]\label{ex:motivating-index}
    Recall that the semigroup $S(V,\nu)$ is generated by the images $\widehat{\nu}(b_i)$ of the Khovanskii basis elements $b_0,\dots,b_{31}\in\mathcal{B}$, computed in Example~\ref{ex:motivating-grobner-valuation}. 
    Also, $G(S)$ is precisely the group generated by these $\widehat{\nu}(b_i)$. 
    Here, $G(S)=\Z^{19}\bigoplus \Z$, meaning that the valuations generate the entire integer lattice as a group. 
    Consequently, $G_0(S)=\Z^{19}\bigoplus \{0\}$. 
    From this, we conclude that $\text{ind}(\rv)=[\Z^d \times \{0\}:G_0(S)]=1$. 
    \end{example}

    \subsection{Newton-Okounkov body}\label{sec:def-newton-okounkov-body}

    The \defword{Newton-Okounkov cone of $(V,\nu)$} 
    is defined to be the closure of the convex hull of the value semigroup $S$ 
    in $\R^{d+1}$. 
    Finally, we define our main computational object: a \emph{Newton-Okounkov body associated to $(V, \nu)$}. 
    
    \begin{definition}[Newton-Okounkov body]\label{def:nob-general}
        Let $(V,\nu)$ be a subspace equipped with a valuation $\nu$ as defined in Section~\ref{sec:valuation}. The \defword{Newton-Okounkov body $\Delta(V,\nu)$ associated to $(V,\nu)$} is defined to be the base of the Newton-Okounkov cone $\text{cone}(V,\nu)$, that is, 
        $\Delta(V,\nu):=\text{cone}(V,\nu)\cap (\R^d\times \{1\}).$ %
        Equivalently, 
        \[\Delta(V,\nu) = \overline{\conv\left(\bigcup\limits_{k>0}^{} \{\nu(f)/k~|f\in V^k s^k\}\right)}\subset \R^d~.\]
    \end{definition}
    
    We conclude this section by returning to our motivating example and illustrating the definitions and notation introduced here.  

    \begin{example}[Motivating Example, continued]\label{ex:motivating-nob-construct}
    In Example~\ref{ex:motivating-grobner-valuation}, we determined a Khovanskii basis~\eqref{eq:wnt-khovanskii-basis} associated to the Wnt network, and from this basis we constructed the associated semigroup $S(V,\nu)$. 
    The cone $\cone(V)$ is the full-dimensional cone in $\R^{20}$ spanned by the columns $a^{(1)},\ldots,a^{(32)}\in \Z^{20}$ of $A$. 
    Writing the column as $a^{(i)}$ as $\left(a_1^{(i)}, a_2^{(i)}, \dots, a_{19}^{(i)}, a_{20}^{(i)}\right)$, the generators of the Newton-Okounkov body $\nobv$ are obtained as follows: 
    divide each of the first $19$ components 
    by the last component 
    $a_{20}^{(i)}$. 
    What remains is $\widetilde{a^{(i)}} := \left(a_{1}^{(i)}/a_{20}^{(i)},~a_2^{(i)}/a_{20}^{(i)},\dots,a_{19}^{(i)}/a_{20}^{(i)}\right) \in \R^{19}$, for $i=1,\dots,32$.     
    The Newton-Okounkov body for Wnt is then the convex hull of the~$\widetilde{a^{(i)}}$ for $i=1,\dots,32$.
    It is a $19$-dimensional polytope sitting in $\R^{19}$. 
    \end{example}

    In Proposition~\ref{prop:kk-nob-bound}, we will see how to use the Newton-Okounkov body associated to $(V,\nu)$ to bound the number of isolated solutions to a general polynomial system in $V$. 
    Kaveh and Khovanskii~\cite{kk-nob-1} showed that, analogous to the setting of a toric variety and its corresponding Newton polytope, asymptotic behavior of the Hilbert function of a graded algebra associated to an arbitrary projective variety can be understood from combinatorial data of its Newton-Okounkov body. 
    As we will see in the next section, a consequence of this is that 
    the volume of the Newton-Okounkov body associated to $(V,\nu)$ is an invariant that bounds the number of effective solutions to a general polynomial system \emph{drawn from} $V$. %

\subsection{The self-intersection index: a bound on isolated solutions}\label{sec:kk-thm}%
    Kushnirenko's theorem implies that the number of solutions (in ${(\C^*)}^d$) to a general polynomial system with support in a finite set $A\subset \Z^d$ is an invariant of the polynomial vector space $V$ generated by the monomials with exponents in $A$. 
    In other words, 
    the number of solutions to a general system of polynomials drawn from $V$ depends \emph{on the subspace} $V$, and has less on the particular polynomial system drawn from it. 
    This generalizes this to more general vector spaces $V$: \cite{kaveh-khovanskii-first-def-biii-2010} showed that 
    the number of solutions (in some distinguished subset of an algebraic variety) of a system of rational functions in a rational space is an invariant of the subspace $V$, and is called the \emph{self-intersection index} of V (Definition~\ref{def:int-index}). 
    Moreover, the actual number of solutions is computed using the constructions from the previous section (Proposition~\ref{prop:kk-nob-bound}).

    We carefully explain the results from~\cite{kaveh-khovanskii-first-def-biii-2010} now. 
    As in the previous section, 
    $X$ is an irreducible $d$-dimensional complex algebraic variety with function field $\mathbb{C}(X)$, 
    and $V$ is a nonzero finite-dimensional subspace of rational functions on $X$.

    \begin{definition}[smooth locus and base locus]\label{def:smooth-loc-base-loc}
    Let $(V_1,\dots,V_d)$ be a $d$-tuple of nonzero finite-dimensional subspaces of rational functions on $X$. 
    Let ${\bf V}:= V_1 \times \dots \times V_d$. 
    Take $\uv \subset X$ to be the set of all nonsingular points in $X$ at which every function in each $V_i$ is regular. 
    Let $\zv\subset \uv$ be the collection of all common zeros from each $V_i$:  
    \[\zv :=\bigcup_{i=1}^d\{x\in U_{\bf V}~|~f(x)=0~\text{~for~all~}f\in V_i\}.~\]
    We refer to $\uv$ as the \defword{smooth locus} 
    of $X$ with respect to%
    ${\bf V}$ and to $\zv$ as the \defword{base locus} of ${\bf V}$. 
    \end{definition}
        
    We will only consider vector spaces $V$ spanned by a finite set of polynomial functions. 
    Consequently, every function in $V$ is regular, and $\uv$ is just the set of nonsingular points of $X$. 
    We will be able to count \emph{effective solutions} to \emph{general} polynomial systems. 
            
    \begin{definition}[effective solution]\label{def:effective-solution}
    Let $(f_1,\dots,f_d)\in V_1\times\dots\times V_d$ be a $d$-tuple of rational functions. 
    An \defword{effective solution} of $f_1=\dots=f_d=0$ is a nondegenerate $x$ in the set $\uv\setminus\zv$ such that $f_1(x)=\dots=f_d(x)=0$. 
    \end{definition}

    Kaveh and Khovanskii~\cite{kaveh-khovanskii-first-def-biii-2010} showed that any $d$-tuple $(V_1,\dots,V_d)$ of rational functions has a proper algebraic subvariety $\textbf{R} \subset \textbf{V}:=V_1\times \dots \times V_d$ such that for any  $(f_1,\dots,f_d)\in \textbf{V}\setminus\textbf{R}$: 
        (i) 
        the number of solutions to the system $f_1=\dots=f_d=0$ in the set $U_{\textbf{V}}\setminus Z_{\textbf{V}}$ is independent of the choice of general $(f_1,\dots,f_d)$, %
        and 
        (ii)  
        each solution $a\in U_{\textbf{V}}\setminus Z_{\textbf{V}}$ of the system $f_1=\dots=f_d=0$ is nondegenerate.         
    We call such a $d$-tuple $(f_1,\dots,f_d)$ of rational functions a \defword{general system} drawn from $V$. 
        
    \begin{definition}[intersection index]\label{def:int-index}
        Let $(V_1,\dots,V_d)$, where $V_i\subset \C(X)$, be a $d$-tuple of subspaces of rational functions on $X$. 
        Let $(f_1,\dots,f_d)\in V_1\times \dots \times V_d$ be a \emph{general system}. 
            The \defword{(birationally-invariant) intersection index $[V_1,\dots,V_d]$} is defined to be the number of 
            effective solutions to  $f_1=\dots=f_d=0$. 
    \end{definition}
    
    In this paper, we will only consider the case where the $V_i$ are all equal, so the $d$-tuple from Definition~\ref{def:int-index} is $(V,\dots,V)$ for some subspace $V$ of rational functions on $X$. 
    In this case, $[V,\dots,V]$ is called the \defword{(self-)intersection index of $V$}.
    We emphasize that the intersection index of $V$ counts effective solutions to a system $f_1=\dots=f_d=0$,  
    where each $f_i$ is a general element 
    drawn from $V$.

    We now introduce the key result central to our application: the self-intersection index of a subspace of rational functions is proportional to the volume of its associated Newton-Okounkov body. 
    \begin{prop}[Kaveh-Khovanskii~\cite{kk-nob-1}]\label{prop:kk-nob-bound}
        Let $X$ be a $d$-dimensional irreducible complex algebraic variety and 
        let $V$ be a nonzero finite-dimensional vector subspace of rational functions on $X$ with Kodaira map $\varphi_V$. 
        If $\varphi_V$ has finite mapping degree, then 
        \[[V,\dots,V]=\dfrac{d!~\deg~\varphi_V}{\ind~\rv} \vold{\Delta(\rv)}~.\]
    \end{prop}

    Proposition~\ref{prop:kk-nob-bound} says that if $\{f_1=\dots=f_d=0\}$ is a system of polynomial equations, where each $f_i$ is general in $V$, 
    then the number of effective solutions to $\{f_1=\dots=f_d=0\}$ 
    can be computed explicitly in terms of the algebraic-geometric objects introduced in this section.  

    \begin{example}[Motivating Example, continued]\label{ex:motivating-biii}
    In Example~\ref{ex:motivating-nob-construct}, 
    we constructed the 19-dimensional Newton-Okounkov body of the Wnt network, a 19-dimensional polytope that is the convex hull of the $\widetilde{a^{(1)}}, \dots, \widetilde{a^{(32)}}$.
    So, $d=19$, and also 
    the normalized volume $19!\vnob$ of this convex body is 32.     
    In Example~\ref{ex:motivating-index}, we computed $\text{ind}(\rv)=1$. 
    In Example~\ref{ex:motivating-Kodaira-map}, we showed that $\deg{\phi_V}=1$. 
    So, by Proposition~\ref{prop:kk-nob-bound}, the self-intersection index of $V$ is 32, from which we conclude  the number of effective solutions to the polynomial system~\eqref{eq:wnt-aug-sys} is at most 32.

    \end{example}

    As described in this section, a Newton-Okounkov body is associated to a 
    polynomial vector space $V$ and a valuation $\nu$ (more specifically, to a finite \emph{Khovanskii basis} $\mathcal{B}$ for $(V,\nu)$) via a graded ring $\rv$. %
    Through our motivating example, we showed how to apply the setup to a chemical reaction network to obtain a Newton-Okounkov body associated to a chemical reaction network, whose volume gives a bound on the maximum number of steady states of the network. 
    In the next section, we state this process more explicitly 
    (Procedure~\ref{proc:nob-compute}). 
    While the motivation in this paper is polynomial systems arising from chemical reaction networks, we emphasize that the procedure applies more generally to any sparse polynomial system one wishes to solve. 

\section{Newton-Okounkov bodies of chemical reaction networks}\label{sec:nob-def-crn}
    Given a chemical reaction network, this section describes how to compute an associated Newton-Okounkov body bound. This first involves finding a vector space of functions that includes the chemical reaction system. After 
    choosing such a  
    vector space $V$, one computes the self-intersection of $V$. 
    The corresponding Newton-Okounkov body bound for the associated network is defined as this self-intersection index.

\subsection{Procedure to compute the self-intersection index of a chemical reaction network}
    We give a procedure for computing the self-intersection index of a chemical reaction system defined by polynomials $\{f_1,\ldots,f_d\}$, where $f_i\in \C[x_1,\dots,x_d]$. 
    Specifically, we construct 
    a vector space $V$ (Section~\ref{sec:vs-graded-alg}), 
    the Kodaira map $\varphi_V$ (Section~\ref{sec:deg-kodaira-map}), 
    a valuation $\nu$ (Section~\ref{sec:valuation}), 
    a Khovanskii basis $\mathcal{B}$ (Section~\ref{sec:khovanskii-basis}), 
    the index of $\rv$ (Section~\ref{sec:ind-graded-algebra}), 
    the Newton-Okounkov body $\nobv$ (Section~\ref{sec:def-newton-okounkov-body}), %
    and finally apply Proposition~\ref{prop:kk-nob-bound}.

    \begin{proc}\label{proc:nob-compute}\ \\
    \emph{Input}: A chemical reaction system defined by polynomials $\{f_1,\ldots,f_d\}$, where $f_i\in \C[x_1,\dots,x_d]$. 
    
    \noindent \emph{Output}:~The self-intersection index for the chemical reaction system, i.e., an upper bound on the number of its positive steady states for any generic choice of rate constants. 
    
    \begin{enumerate}\setcounter{enumi}{-1} 
        \item\label{proc:step-mon-order} Fix a monomial term order $<$.  
        \item\label{proc:step-vs-basis} Choose a generating set $\{v_1,\dots,v_n\}$ for a polynomial vector space $V$ such that each $f_i$ is 
        in the complex span of the $v_j$, i.e. $f_i\in V:=\spn_{\C}\{v_1,\ldots,v_n\}$ for $i = 1, \dots, d$.

        \item\label{proc:step-deg} Compute $\deg(\varphi_V)$ according to Section~\ref{sec:deg-kodaira-map}. 
        \item\label{proc:step-khov-basis} Compute
        a finite
        Khovanskii basis $\mathcal{B}$ for the subalgebra $\rv$ associated to $V=\spn_{\C}\{v_1,\ldots,v_n\}$ according to Section~\ref{sec:khovanskii-basis}. 
        \emph{If a finite Khovanskii basis does not exist, return to Step 0, and choose a different monomial term order $<$.}
        
        \item\label{proc:step-index} Compute $\ind(\rv)$ according to Section~\ref{sec:ind-graded-algebra}. 

        \item\label{proc:step-vol-nob} Compute the volume $\vnob$ of the Newton-Okounkov body $\nobv:=\Delta(V,<,\mathcal{B})$ 
        according to Section~\ref{sec:def-newton-okounkov-body}.     
        \item\label{proc:step-biii} Output the birationally invariant self-intersection index $[V,\dots,V]$, according to Proposition~\ref{prop:kk-nob-bound}. 

        \end{enumerate}
        \end{proc}

    \subsection{Practical Considerations}
    
    \begin{remark}[choosing a basis]
    One strategy for choosing a generating set for input to {\tt SubalgebraBases.m2} is to look for commonalities among the defining polynomials that we wish to solve. 
    For instance, when the defining polynomials of the system admit common factors (c.f.,~a \emph{product decomposition} as in~\cite{msw-product-decomp-bound}) or common summands, we choose these shared polynomials in the list of vector space generators. 
    We apply this strategy in~Example~\ref{ex:pm} and Example~\ref{ex:polly-paper-ex-4.1}.         
    Step~\ref{proc:step-vs-basis} is the most challenging step of Procedure~\ref{proc:nob-compute} and requires some trial-and-error by the user. 
    \end{remark}

    \begin{remark}\label{rmk:proc-computes-effective-solutions}
    The bound $[V,\dots,V]$ computes all \emph{effective} solutions (Definition~\ref{def:effective-solution}) to any general system $g_1=0,\dots,g_d=0$, where $g_i$ is general in $V$. 
    Recall that this means that all solutions counted are outside $Z_{\textbf{V}}$, the base locus of $V$ as in Definition~\ref{def:smooth-loc-base-loc}.    
    Note in particular that this means there could be additional real positive steady states in the base locus. 
    For each example we consider, the base locus is easily computed 
    and does not contain any positive real solutions.%
    \end{remark}
    
    \begin{remark}[grevlex]\label{rmk:grevlex}
    For our examples, in Step~0 of Procedure~\ref{proc:nob-compute}, we will always select the monomial term order grevlex first. 
    The leading term under grevlex yields the smallest exponent under the Gr\"obner valuation (Definition~\ref{def:grobner-valuation}). 
    \end{remark}
    
    \begin{remark}[finite Khovanskii basis]
        A finite Khovanskii basis is not guaranteed. 
        However, if the semigroup $S(V,\nu)$ is finitely generated, then there is a finite Khovanskii basis for $(V,\nu)$~\cite{kaveh-manon}. 
        If a finite Khovanskii basis is not achieved for a chosen valuation, one strategy is to choose a different basis 
        as input to  {\tt SubalgebraBases.m2}. 
        Another strategy is to try a different valuation, like the Gr\"obner valuation induced by a different monomial term order.  
        For Example~\ref{ex:pm}, the same vector space for $V$ basis together with the Gr\"obner valuation corresponding to the {\tt lex} monomial order instead produced an infinite sagbi basis. 
    \end{remark}

    \begin{remark}
    For small examples, the degree of the Kodaira map can be computed by hand. 
    We verify the computation using the Macaulay2 package {\tt Cremona.m2}~\cite{cremona}. 
    Note that {\tt Cremona.m2} requires a homogeneous map as input: in our computations, we multiply each $v_i$ by a power of a homogenizing variable.
    \end{remark}
    
\section{Examples}\label{sec:nob-examples}

    In this section, we explicitly compute Newton-Okounkov bodies of select chemical reaction networks. 
    We also calculate the mixed volume bound for each reaction network, as in~\cite{OSTT,mv-small-networks,NO-dissertation}.
    The supporting computations for each example are archived at:~\url{https://github.com/EliseAWalker/NOBodies-for-ChemicalRxns}. 
    Recall (Remark~\ref{rmk:grevlex}) that for each example, we choose the Gr\"obner valuation induced by the grevlex monomial order when applying Step~\ref{proc:step-mon-order}.

    \begin{example}[Wnt, Motivating Example~\ref{ex:motivating-crn-sec}, concluded]\label{ex:motivating-summary}
    As shown in Example~\ref{ex:motivating-biii}, the Newton-Okounkov body bound for the Wnt network is 32, which is tighter than the mixed-volume bound of 56. 
    Furthermore, there are no positive real solutions in the base locus (indeed, the base locus is empty), and hence we expect no more than 32 positive steady states for Wnt, for generic parameters.
    \end{example}%
    \vspace*{-10pt}
    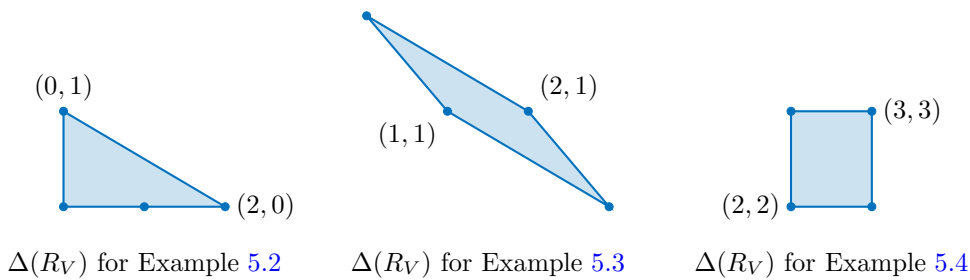
\begin{figure}[ht!]
    \centering
    \begin{tikzpicture}
\begin{axis}[%
width=5.500in,
height=1.600in,
scale only axis,
xmin=0,
xmax=13,
xtick = \empty,
ymin=-1.0,
ymax=2.2,
ytick = \empty,
axis line style={draw=none} 
]

\addplot [color=mycolor1,solid,thick, fill opacity = 0.2, fill = mycolor1,forget plot]
  table[row sep=crcr]{%
1	0\\
3	0\\
1	1\\
1	0\\
};
\addplot[only marks,mark=*,mark size=1.5pt,mycolor1
        ]  coordinates {
    (1,0) (3,0) (1,1) (2,0) 
};
\node at (axis cs:1,1.25) {$(0,1)$};
\node at (axis cs:3.5,0) {$(2,0)$};
\node at (axis cs:2,-0.6) {$\Delta(\rv)$ for Example~\ref{ex:from-ideas-of-nobs}};

\addplot [color=mycolor1,solid,thick, fill opacity = 0.2, fill = mycolor1,forget plot]
  table[row sep=crcr]{%
4.75	2\\
6.75	1\\
7.75	0\\
5.75	1\\
4.75 2\\
};
\addplot[only marks,mark=*,mark size=1.5pt,mycolor1
        ]  coordinates {
    (4.75,2) (6.75,1) (7.75,0) (5.75,1) 
};
\node at (axis cs:5.25,0.75) {$(1,1)$};
\node at (axis cs:7.25,1.25) {$(2,1)$};
\node at (axis cs:6.25,-0.6) {$\Delta(\rv)$ for Example~\ref{ex:pm}};

\addplot [color=mycolor1,solid,thick, fill opacity = 0.2, fill = mycolor1,forget plot]
  table[row sep=crcr]{%
10	1\\
11	1\\
11	0\\
10	0\\
10	1\\
};
\addplot[only marks,mark=*,mark size=1.5pt,mycolor1
        ]  coordinates {
    (10,1) (11,1) (10,0) (11,0) 
};
\node at (axis cs:11.5,1) {$(3,3)$};
\node at (axis cs:9.5,0) {$(2,2)$};
\node at (axis cs:10.5,-0.6) {$\Delta(\rv)$ for Example~\ref{ex:polly-paper-ex-4.1}};

\end{axis}
\end{tikzpicture}
    \vspace*{-10pt}
    \caption{Newton-Okounkov bodies for Examples~\ref{ex:from-ideas-of-nobs},~\ref{ex:pm}, and~\ref{ex:polly-paper-ex-4.1}, respectively.}
    \end{figure}

    \begin{example}[KST~\cite{ideas-nobs}]\label{ex:from-ideas-of-nobs}
    Consider the following system:     
    \begin{equation}\label{eq:poly-sys-2circ-ex-non-gen-coeff}
        \begin{aligned}
    f_1 &= k_1(x^2+y^2) -k_2 x + k_3\\
    f_2 &= k_1(x^2+y^2) -k_4 y + k_5 ~.
    \end{aligned}
    \end{equation}
    \noindent where each $k_i\in\R_{> 0}$. 
    This polynomial system can be 
    obtained from a chemical reaction network with $f_{\ell}=\dot{x}_{\ell}$ since each defining polynomial is of the shape $f_{\ell} = p_{\ell} - x_{\ell} q_{\ell}$ for real polynomials $p_{\ell},~q_{\ell}$ with nonnegative coefficients~\cite{hars-toth-inverse-shape-of-crn-polys}. %
    We note here that some information is known about the 
    coefficients: 
    in particular, the coefficients of $(x^2+y^2)$ in 
    $f_1$ and $f_2$ 
    are equal. 
    
    In this small example, we can explicitly compute the maximum number of complex solutions.
    Geometrically, the system defines the intersection of two circles, so the system has at most 2 isolated complex solutions. 
    Moreover, we produce a witness for two positive, real solutions by taking $k_1=1,~k_2=8,~k_3=7,~k_4=6,~k_5=3$. 
    In other words, the maximum number of steady states for the associated chemical reaction system is 2.
    Next we apply Procedure~\ref{proc:nob-compute}, and show that the associated self-intersection index gives a tight bound. 
    
    \textit{Input:} The chemical reaction system $\{f_1,f_2\}\subset\C[x,y]$. 
    
    \textbf{Step~\ref{proc:step-vs-basis}:}~Let $V=\spn_{\C}\{1,x,y,(x^2+y^2)\}$. 
    Here, since the support of $f_1$ and $f_2$ share the summand $(x^2+y^2)$, we choose the polynomial instead of the individual monomials $x^2$ and $y^2$ as generators of the vector space. 

    \textbf{Step~\ref{proc:step-deg}:}~The Kodaira map of $V$ is defined by $(x,y)\mapsto [s:sz:ys:(x^2+y^2)s]$, and it has degree 1.

    \textbf{Step~\ref{proc:step-khov-basis}:}~Let $Vs$ be generated by $\{s,xs,ys,(x^2+y^2)s\}$. 
    We verify that the generating set at height 1 is in fact a Khovanskii basis 
    for the graded algebra $\rv$ associated to $V$: so, $\mathcal{B}=\{s,xs,ys,(x^2+y^2)s\}$.

    \textbf{Step~\ref{proc:step-index}:}~The valuations of the elements of $\mathcal{B}$ are $(0,0,1), (1,0,1), (0,1,1),$ and $(2,0,1)$. 
    They generate the value semigroup $S(V,\nu)$. 
    The subgroup $G_0(\rv) = \Z^2\oplus \{0\}$. 
    The index of the lattice $\ind~\rv$, that is, the index of $G_0(\rv)$ in $\Z^2\oplus \{0\}$, is 1. 
    
    \textbf{Step~\ref{proc:step-vol-nob}:}~Then $\nobv$ is the triangle $\conv\{(0,0),(0,1),(2,0)\}$ in $\R^2$.

    \textbf{Step~\ref{proc:step-biii}:}~The self-intersection index $[V,V]$ is $2!\vnob = 2$. 
    
    \textit{Output}: This chemical reaction system has a maximum of 2 positive steady states (outside the base locus). 
    Note that for this example, the base locus, which consists of all solutions to $\{1=0, x=0, y=0, (x^2+y^2)=0\}$, is empty. %

    A general system drawn from $V$ has the form 
    \[
    \begin{cases}
    \kappa_1(x^2+y^2) + \kappa_2 y + \kappa_3 x + \kappa_4 = 0\\
    \kappa_5(x^2+y^2) + \kappa_6 y + \kappa_7 x + \kappa_8 = 0~.
    \end{cases}
    \]
    \noindent with $\kappa_i\in \C$. Thus, the mixed volume of a general system in $V$ is 4. 
    However, any such generic system has exactly 2 solutions, since, the system defines the intersection of two circles. 
    In other words, the maximum number of complex-number steady states is equal to the Newton-Okounkov body bound. 
    This Newton-Okounkov body bound is also tighter than the mixed volume bound of 4 for the original system~\eqref{eq:poly-sys-2circ-ex-non-gen-coeff}. %

    \end{example}

    \begin{example}[PM~\cite{perez-millan-thesis}]\label{ex:pm}
    The following system of polynomials comes from a chemical reaction network and was studied as a network with \textit{absolute concentration robustness} in~\cite[Example 6.5.3]{perez-millan-thesis}: %
    \begin{equation}\label{eq:pm}
        \begin{aligned}
        f_1(x) &= x_1[(x_1-1)^2 + (x_2-2)^2] \\
        f_2(x) &= x_2[(x_1-1)^2 + (x_2-2)^2]
        \end{aligned}
    \end{equation}
    It is straightforward to see that this system has exactly one positive steady state $(x_1^*,x_2^*)=(1,2)$. 
    In what follows, we will ignore the fact that this system is simple to solve by inspection and highlight how to apply our tools from Section~\ref{sec:nob-def-crn} to this toy example. 
    In particular, we 
    emphasize that the degree of the Kodaira map is not always 1 and needs to be computed explicitly in order to apply Proposition~\ref{prop:kk-nob-bound}.

    \textit{Input}: the chemical reaction system $\{f_1,f_2\}\subset\C[x_1,x_2]$. 
    
    \textbf{Step~\ref{proc:step-vs-basis}:}~This system belongs to the vector space $V = \spn_{\C}\{x_1(x_1^2+x_2^2 -2x_1),~x_1(5-4x_2),~x_2(x_1^2+x_2^2 -2x_1),~x_2(5-4x_2)\}$. 
    In particular, $f_1$ is equal to the sum of the first two basis elements of $V$, and $f_2$ is the sum of the last two basis elements.

    \textbf{Step~\ref{proc:step-deg}:}~The Kodaira map of $V$ is defined by $(x,y)\mapsto [x_1(x_1^2+x_2^2 -2x_1)s: x_1(5-4x_2)s: x_2(x_1^2+x_2^2 -2x_1)s: x_2(5-4x_2)s]$, and it has degree 2. 

    \textbf{Step~\ref{proc:step-khov-basis}:}~The generating set induces the Khovanskii basis $\mathcal{B}=\{x_1(x_1^2+x_2^2 -2x_1)s,~x_1(5-4x_2)s,~x_2(x_1^2+x_2^2 -2x_1)s,~x_2(5-4x_2)s\}$
    for the graded algebra $\rv$ associated to $V$.

    \textbf{Step~\ref{proc:step-index}:}~%
    In Example~\ref{ex:pm-index-g0}, we computed that $\ind(\rv)=[\Z^2\oplus\{0\}:G_0(\rv)]=1$.%
    
    \textbf{Step~\ref{proc:step-vol-nob}:}~The Newton-Okounkov body $\nobv$ has vertices $(3,0),(1,1),(2,1),(0,2)$ and volume 1. 

    \textbf{Step~\ref{proc:step-biii}:}~The self-intersection index $[V,V]$ is $d!\cdot \deg~\varphi_L\cdot\vnob /~\ind~\rv= 2!\cdot 2\cdot 1 / 1= 4$. 
    
    \textit{Output}: The maximum number of positive steady states (outside the base locus) is 4. 
    
    Moreover, for this example, no positive solutions are missed by the self-intersection bound. 
    Indeed, the base locus of $V$ contains three points: $Z_{{\bf V}}= \{ (1\pm.75i, 1.25), (0,0)\}$.  
    While four is an over-calculation for the single known positive steady-state, the steady-state $(1,2)$ actually has multiplicity four, and so the calculation for the number of roots (counted with multiplicity) is tight.

    \end{example}

    \begin{example}[BCY~\cite{BorosCraciunYu-InfinitePosSS}]\label{ex:polly-paper-ex-4.1}
        Networks with infinitely many positive steady states in a compatibility class were investigated in \cite{BorosCraciunYu-InfinitePosSS}. 
        These networks were the first such ``weakly reversible'' networks discovered, and \cite{BorosCraciunYu-InfinitePosSS} showed that these networks 
        have infinitely many positive steady states defined by a curve of steady states arising from a common factor of the individual equations in the chemical reaction system. 
        However, in general, the number of \emph{isolated} (complex) solutions to these networks is small.     
    \begin{figure}[ht!]
        \centering
    
                \begin{tikzpicture}[scale=0.8]
            \node[] (0) at (-4,0) {\small{$\ce{0}$}};
            \node[] (x) at (-1.5,0) {\small{$\ce{X}$}};
            
            \node[] (y) at (-4,2) {\small{$\ce{Y}$}};
            \node[] (xy) at (-1.5,2) {\small{$\ce{X + Y}$}};
    
            \draw[->,> = latex]  (0) -- node[anchor=north] {\scriptsize{$1$}} (x);%
            \draw[->,> = latex]  (xy) -- node[anchor=south] {\scriptsize{$5$}} (y);
    
            \draw[->,> = latex]  (xy) -- node[anchor=east] {\scriptsize{$1$}} (x);
    
            \draw[->,> = latex]  (y) -- node[anchor=east] {\scriptsize{$1$}} (0);
    
            \node[] (2x) at (5,0) {\small{$\ce{2X}$}};
            \node[] (3x) at (7.5,0) {\small{$\ce{3X}$}};
            
            \node[] (2xy) at (5,2) {\small{$\ce{2X + Y}$}};
            \node[] (3xy) at (7.5,2) {\small{$\ce{3X + Y}$}};
    
            \draw[->,> = latex]  (2x) -- node[anchor=north] {\scriptsize{$1$}} (3x);%
            \draw[->,> = latex]  (3xy) -- node[anchor=south] {\scriptsize{$1$}} (2xy);
    
            \draw[->,> = latex]  (3x) -- node[anchor=east] {\scriptsize{$1$}} (3xy);
    
            \draw[->,> = latex]  (2xy) -- node[anchor=east] {\scriptsize{$5$}} (2x);
            
            \node[] (2y) at (0.25,0) {\small{$\ce{2Y}$}};
            \node[] (x2y) at (2.75,0) {\small{$\ce{X + 2Y}$}};
            
            \node[] (3y) at (0.25,2) {\small{$\ce{3Y}$}};
            \node[] (x3y) at (2.75,2) {\small{$\ce{X + 3Y}$}};
    
            \draw[->,> = latex]  (3y) -- node[anchor=east] {\scriptsize{$1$}} (2y);%
            \draw[->,> = latex]  (x3y) -- node[anchor=south] {\scriptsize{$1$}} (3y);
    
            \draw[->,> = latex]  (x2y) -- node[anchor=east] {\scriptsize{$5$}} (x3y);
    
            \draw[->,> = latex]  (2y) -- node[anchor=north] {\scriptsize{$1$}} (x2y);
            
            \node[] (2x2y) at (10,0) {\small{$\ce{2X + 2Y}$}};
            \node[] (3x2y) at (13,0) {\small{$\ce{3X + 2Y}$}};
            
            \node[] (2x3y) at (10,2) {\small{$\ce{2X + 3Y}$}};
            \node[] (3x3y) at (13,2) {\small{$\ce{3X + 3Y}$}};
    
            \draw[->,> = latex]  (2x2y) -- node[anchor=north] {\scriptsize{$5$}} (3x2y);%
            \draw[->,> = latex]  (3x3y) -- node[anchor=south] {\scriptsize{$1$}} (2x3y);
    
            \draw[->,> = latex]  (3x2y) -- node[anchor=east] {\scriptsize{$1$}} (3x3y);
    
            \draw[->,> = latex]  (2x3y) -- node[anchor=east] {\scriptsize{$1$}} (2x2y);
            \end{tikzpicture}
    
        \caption{A chemical reaction network with infinitely many positive steady states, but only one isolated positive steady state.}
        \label{fig:polly-ex-network}
    \end{figure}
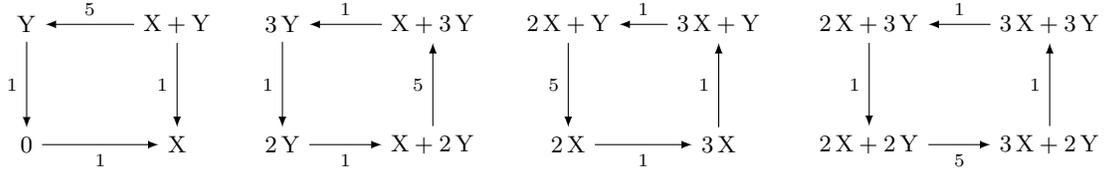
    
        Consider the chemical reaction network in Figure~\ref{fig:polly-ex-network} from \cite[Example 4.1]{BorosCraciunYu-InfinitePosSS}. 
        The corresponding chemical reaction system is given by: 
    \[
    \begin{cases}
        \dot{x} = (x^2 y^2 + x^2 + y^2 + 1 - 5xy)[1 - xy]\\
        \dot{y} = (x^2 y^2 + x^2 + y^2 + 1 - 5xy)[x - y]~.
    \end{cases}
    \]
        The set of positive steady states consists of the non-isolated points $(x,y)$ on the curve $x^2 y^2 + x^2 + y^2 + 1 - 5xy = 0$, and the only isolated solution that is positive is the steady state $(1,1)$~\cite{BorosCraciunYu-InfinitePosSS}. 
        Notice that there is one more complex (non-positive) solution to $\dot{x}=\dot{y}=0$, namely $(-1,-1)$.
        
        \textbf{Step~\ref{proc:step-vs-basis}:}~Let $p_1:=x^2 y^2 + x^2 + y^2 + 1 - 5xy$ (the common factor of the system's two defining polynomials), and let $p_2:=1$, $p_3:=xy$, $p_4:=x$, $p_5:=y$ (these are the monomials in the remaining factors). 
        Notice that $\dot{x} = p_1(p_2 - p_3)$ and $\dot{y} = p_1(p_4 - p_5)$.
        Take $V=\spn_{\C}\{p_1p_2,~p_1p_3,~p_1p_4,~p_1p_5\}$, so $\dot{x}, \dot{y} \in V$.

        \textbf{Step~\ref{proc:step-deg}:}~The Kodaira map of $V$, defined by $(x,y)\mapsto [p_1p_2s:~p_1p_3s:~p_1p_4s:~p_1p_5s]$, has degree 1.

        \textbf{Step~\ref{proc:step-khov-basis}:}~The generating set induces a Khovanskii basis $\mathcal{B}=\{p_1p_2s,~p_1p_3s,~p_1p_4s,~p_1p_5s\}$ for the associated graded algebra $\rv$.

        \textbf{Step~\ref{proc:step-index}:}~The value semigroup is generated by the valuations of the Khovanskii basis elements, namely: $(2,2,1)$, $(2,3,1)$, $(3,2,1)$, and $(3,3,1)$. 
        It is straightforward to see that $(0,1,0) = (2,3,1) - (2,2,1)$ and $(1,0,0)=(3,2,1)-(2,2,1)$ are in $G_0(S)$ and generate $\Z^2\oplus \{0\}$. 
        So, $\ind(\rv)$ is 1: 
        
        \textbf{Step~\ref{proc:step-vol-nob}:}~Then $\nobv$ is the square $\conv\{(2,2),~(2,3),~(3,2),~(3,3)\}$ in $\R^d$, with volume 1.
    
        \textbf{Step~\ref{proc:step-biii}:}~The self-intersection index $[V,\dots,V]$ is $d!\deg(\varphi_V) \vnob /~\ind{\rv}= 2!\cdot 1 \cdot 1 / 1 = 2$. 
    
        \textit{Output}: The maximum number of positive steady states (outside the base locus) is at most 2.

        The base locus consists only of purely complex solutions, hence no positive solutions are missed by the Newton-Okounkov body bound. 
        
        In contrast, by Definition~\ref{def:mv-crn}, the mixed volume of the chemical reaction system is 18. 
        We already saw that the maximum number of isolated positive steady states is 1. 
        This is a great example of the volume of the self-intersection index achieving a tighter bound than the mixed volume: the mixed-volume overcount is $18 - 1 = 17$, but the analogous \emph{Newton-Okounkov overcount} is $2 - 1 = 1$.%

    \end{example}

    Table~\ref{tab:bounds-for-all-examples} summarizes the bounds for each of the networks considered in this section. We observe that the Newton-Okounkov bound is always better (or at least as good) as the mixed-volume bound.  %

        \begin{table}[ht!]
            \centering
            \begin{tabular}{c|c|c|c|c|c}
            nickname, & example %
               & max number  & max number  & mixed-volume & Newton-Okounkov \\
                {[}source{]} & & steady states &  $\C$-steady-states &  bound &  body bound \\    
            \hline
            Wnt~\cite{wnt} & Example~\ref{ex:motivating-summary} & $\geq 3$ & 9 & 56 & 32\\ 
            KST~\cite{ideas-nobs} & Example~\ref{ex:from-ideas-of-nobs} %
                & 2 & 4 & 4 & 2\\   
            PM~\cite{perez-millan-thesis} & Example~\ref{ex:pm} %
                & 1 & 4 & 4 & 4\\ 
            BCY~\cite{BorosCraciunYu-InfinitePosSS} & Example~\ref{ex:polly-paper-ex-4.1} %
                & 1 & 2 & 18 & 2
            \end{tabular}
            \caption{The various root counts associated to the chemical reaction systems in Section~\ref{sec:nob-examples}. The Newton-Okounkov body bound matches or outperforms the mixed-volume bound for every example. }
            \label{tab:bounds-for-all-examples}
        \end{table}

\section{Discussion}

The self-intersection index of a finite dimensional subspace of rational functions $V$ bounds the number of isolated 
solutions to a general polynomial system in the linear span of a Khovanskii basis for the associated graded algebra $\rv$. %
However, in practice, very few examples of Newton-Okounkov bodies had 
been computed. 
In this paper, %
we defined the first notion of a Newton-Okounkov body of a chemical reaction system, and gave a procedure for computing this body and the resulting intersection index  (Procedure~\ref{proc:nob-compute}). 
Applying this procedure on 
concrete examples, we showed that the volume of the intersection index of a chemical reaction network 
improves 
the mixed volume bound (Section~\ref{sec:nob-examples}). 
These examples provide a proof-of-concept for our new method. 
In summary, the Newton-Okounkov body is a promising new tool for achieving our ultimate goal of tight bounds on the number of steady states of an arbitrary reaction network. 
We conclude with some final remarks and ideas for next steps.

    \subsection{When can a Newton-Okounkov body improve bounds?}\label{sec:nob-improves-bounds}
    There are two important situations when a Newton-Okounkov body achieves better upper bounds on the maximum number of steady states of chemical reaction networks than those discussed in Section~\ref{sec:classical-bounds}. 
    
    First, we recall that the Kushnirenko Theorem and the Bernstein Theorem take into account only the monomial support and are independent of precise rate constants in the chemical reaction system~\eqref{consys} (provided that the rate constants are sufficiently general). 
    However, in certain applications or for specific networks, the rate constants are likely to be non-generic. 
    For instance, we may 
    \emph{a priori} know some relationship among the rate constants, such as equality. 
    On the other hand, we may have specific values for some or all of the rate constants. 
    For example, the chemical reaction system from Example~\ref{ex:polly-paper-ex-4.1} has rate constants $\kappa =
    (k,k,5k,k,k,5k,k,k,k,k,k,k,k,k,5k,k,5k,k,k,k)$, for $k\in\R_{\geq 0}$. 
    Recall that for this example, the Newton-Okounkov body bound greatly outperformed the mixed-volume bound. 
    We anticipate the theory of Newton-Okounkov bodies for chemical reaction networks will improve bounds on the maximum number of steady states in similar cases where we can utilize extra information about the rate constants. 
    
    Second, the intersection index of a subspace $V$ counts solutions to a polynomial system drawn from $V$ outside an algebraic variety 
    that 
    contains the base locus and the set of poles of $V$. 
    The mixed volume on the other hand counts \textit{all} isolated solutions in $(\C^*)^d$. 
    Hence, using the theory of Newton-Okounkov bodies, we can restrict to counting solutions in some specified Zariski-open subset of $(\C^*)^d$, for example. 
    With an appropriately chosen basis for $V$, 
    the theory of Newton-Okounkov bodies for chemical reaction networks can bound steady states outside the specified base locus. 

    \subsection{Non-conservative networks}
    The examples in Section~\ref{sec:nob-examples} were all pathological in the sense that none of the networks were conservative. 
    Recall that for a conservative network, each species occurs in at least one nonnegative conservation law. 
    Based on our investigations, we conjecture the following. 
    \begin{conjecture}
    For every conservative reaction network, the mixed-volume bound is equal to the Newton-Okounkov bound. 
    In other words, conservative networks are ``Bernstein-general''. 
    \end{conjecture}

    \subsection{Non-polynomial chemical reaction systems}
    We remark that the theory developed in Section~\ref{sec:nob-setup}  supports systems of the form $p_1/q_1=0,\dots,p_d/q_d=0$, where $p_i,q_i$ are polynomials in $\C[x_1,\dots,x_d]$. 
    That is, the self-intersection index is defined for any nonzero finite dimensional vector space $V$ of \emph{rational functions} on a projective variety $X$, and hence 
    bounds the number of effective solutions to systems of \emph{rational} equations. 
    We considered here chemical reaction systems arising from mass-action kinetics. 
    Other kinetic assumptions, or other steady-state approximations, produce alternative chemical reaction systems that are not polynomials. 
    \begin{question}
    How tight are Newton-Okounkov body bounds associated to rational steady-state approximations, e.g., Michaelis–Menten kinetics, or associated to positive rational steady-state parametrizations (c.f.,~\cite[Section 6.6.4]{NO-dissertation})? 
    \end{question}

    \subsection{Best bounds}\label{sec:best-basis}
    Procedure~\ref{proc:nob-compute} detailed a method for computing \emph{one possible} Newton-Okounkov body for a chemical reaction system. 
    However, the resulting Newton-Okounkov body is not well-defined. 
    There are a number of choices, including choice of valuation/monomial term order and choice of a vector space basis, made in Procedure~\ref{proc:nob-compute} that can potentially yield different associated Newton-Okounkov bodies. 
    For instance, the defining polynomial system in~\eqref{consys} may in principle give different vector spaces $V$ depending on the order of the variables. 
    We would like one well-defined definition of a Newton-Okounkov body, or at least some detailed guiding principles for how best to define one to achieve the tightest bound. 
    More explicitly, a next research step is to remove the ``trial-and-error'' in applying the Procedure~\ref{proc:nob-compute}; see Remark~\ref{rmk:choosing-vs-basis}). 
    
    \begin{question}\label{ques:choose-right-basis}
    Under which hypotheses on a basis for $V$ does the associated Newton-Okounkov body bound give the tightest bound on the maximum number of steady states of a chemical reaction system? 
    \end{question}
    
    In this work, we showed how Newton-Okounkov body theory can and should be used in practice to obtain good bounds on the number of solutions to polynomial systems arising from applications. 
    Moving forward, we are hopeful that Procedure~\ref{proc:nob-compute} will be an important computational tool for applied mathematicians. 
    A researcher interested in counting solutions to their own polynomial system can make simple modifications of our code and subsequently calculate the associated self-intersection index, thereby obtaining better bounds than was previously possible.

\subsection*{Acknowledgements}
    Ideas leading to this research emerged during the Newton-Okounkov Bodies Working Seminar organized by Frank Sottile and Emanuele Ventura at Texas A\&M University in Spring 2018. 
    When this research was initiated, 
    NO was supported by an American Fellowship from the AAUW. 
    We thank 
    Michael Burr, 
    Heather Harrington, 
    Anne Shiu,  
    Aleksandra Sobieska,  
    Frank Sottile, 
    and 
    Ruth Williams 
    for helpful discussions.

\newpage

\bibliography{nobs.bib}

\appendix

\section{Notation}\label{app:notation}
We collect the notation used throughout the paper. 

        \begin{itemize}
            \item $\N$, the natural numbers, including 0
            \item $e_i\in \R^d$, the $i$-th standard basis vector
            \item $F^*:=F\setminus \{0\}$, the set of nonzero elements of $F$ 
            \item $X$, an irreducible $d$-dimensional complex algebraic variety
            \item $\C(X)$, function field of $X$
            \item $V\subset \C(X)$, a nonzero finite-dimensional complex vector space of rational functions on $X$
            \item $s$, a formal grading variable
            \item $V^k\subset \C(X)$, the subspace spanned by all $k$-fold products of elements in $V$
            \item %
            $\rv:=\displaystyle \bigoplus_{k\geq 0} V^ks^k$, the graded subalgebra of $\C(X)[s]$ associated to $V$ 
            \item $\nu$, a valuation on $\C(X)$            
            \item $\widehat{\nu}$, a ``lifted'' valuation on $\rv$
            \item $S:=S(V,\nu)=\widehat{\nu}(\rv)\subset \Z^d\oplus \Z$, the value semigroup associated to $(V,\nu)$

            \item $\mathcal{B}\subset \rv$, a finite Khovanskii basis for $\rv$
            \item $G(\rv)=\Z S(V,\nu)$, the group generated by $S(V,\nu)$
            \item $\pi : \R^{d+1} \rightarrow \R$ given by projection onto the last coordinate, i.e., $(x_1, \dots, x_d, x_{d+1})\mapsto x_{d+1}$
            \item $G_0(\rv) := \pi^{-1}(0) \cap G(\rv)$, the subgroup of $\Z^n\bigoplus \{0\}$ generated by $S(V,\nu)$              
            \item $\text{ind}(\rv)$, the index of $G_0(\rv)$ as a subgroup of $\Z^d \times \{0\}$
            \item $\Delta(V,\nu,
            \mathcal{B})$, the Newton-Okounkov body associated to $V$, given valuation $\nu$ and Khovanskii basis $\mathcal{B}$            
        \end{itemize}

\end{document}